\documentclass[aps,pra,twocolumn,floatfix]{revtex4-1}
\usepackage{amssymb,bm,dcolumn,amsmath,color,graphicx,graphics}

\newcommand{\eref}[1]{Eq.~(\ref{#1})}
\newcommand{\tref}[1]{Table~\ref{#1}}

\begin{document}

\title{Optical clocks based on the Cf$^{15+}$ and Cf$^{17+}$ ions}
\date{\today}

\begin{abstract}
Recent experimental progress in cooling, trapping, and quantum logic spectroscopy of highly-charged ions (HCIs)  made HCIs accessible for high resolution
spectroscopy and precision fundamental studies. Based on these achievements, we explore a possibility to develop optical clocks using transitions between the ground and a low-lying excited state in the Cf$^{15+}$ and Cf$^{17+}$ ions. Using a high-accuracy relativistic method of calculation we predicted the wavelengths of clock transitions, calculated relevant atomic properties, and analyzed a number of systematic effects (such as the electric quadrupole-, micromotion-, and quadratic Zeeman shifts of the clock transitions) that affect the accuracy and stability of the optical clocks. We also calculated magnetic dipole hyperfine-structure constants of the clock states and the blackbody radiation shifts of the clock transitions.
\end{abstract}

\author{S. G. Porsev$^{1,2}$}
\author{U. I. Safronova$^{3}$}
\author{M. S. Safronova$^{1,4}$}
\author{P. O. Schmidt$^{5,6}$}
\author{A. I. Bondarev$^{2,7}$}
\author{M. G. Kozlov$^{2,8}$}
\author{I. I. Tupitsyn$^{2,9}$}

\affiliation{
$^1\!$Department of Physics and Astronomy, University of Delaware, Newark, Delaware 19716, USA\\
$^2\!$Petersburg Nuclear Physics Institute of NRC ``Kurchatov Institute'', Gatchina, Leningrad District, 188300, Russia,\\
$^3\!$Physics Department, University of Nevada, Reno, Nevada 89557, USA\\
$^4\!$Joint Quantum Institute, National Institute of Standards and Technology and the University of Maryland, College Park, Maryland 20742, USA\\
$^5\!$Physikalisch-Technische Bundesanstalt, Bundesallee 100, 38116 Braunschweig, Germany \\
$^6\!$Institut f\"ur Quantenoptik, Leibniz Universit\"at Hannover, Welfengarten 1, 30167 Hannover, Germany\\
$^7\!$Center for Advanced Studies, Peter the Great St. Petersburg Polytechnic University, Polytechnicheskaya 29, St. Petersburg, 195251, Russia \\
$^8\!$St. Petersburg Electrotechnical University LETI, Prof. Popov Str. 5, St. Petersburg, 197376, Russia\\
$^9\!$Department of Physics, St. Petersburg State University, Ulianovskaya 1, Petrodvorets, St. Petersburg, 198504, Russia}

\maketitle
\section{Introduction}
\label{Intro}
Recent years marked a rapid development of both highly-charged ion (HCI) theory and experiment.
An experimental progress in cooling and trapping of HCIs using sympathetic cooling made them accessible for high resolution
spectroscopy and precision fundamental studies \cite{SchVerSch15,KozSafCre18,MicLeoKin20}.

The pioneering works of Schiller~\cite{schiller_hydrogenlike_2007} and Berengut {\it et al.}~\cite{BerDzuFla10} proposed to use optical transitions in HCIs for frequency metrology and tests for a variation of the fundamental constants. In a number of subsequent theoretical studies (see recent review~\cite{KozSafCre18} and references therein) it was demonstrated that a number of HCIs have narrow transitions lying in the optical frequency range, which can be used for developing high-accuracy clocks as well as other properties desirable for precision frequency metrology.

In comparison to neutral atoms, HCIs have several advantages. They have a more compact size and, hence, are less sensitive to external electric field perturbations.
Preliminary estimates of a systematic uncertainty that can be obtained using shift mitigation and cancellation strategies suggest that the uncertainties well below $10^{-18}$ may be achievable~\cite{DerDzuFla12,DzuDerFla12,DzuDerFla13}. The sensitivity of an HCI clock transition to a variation of the fine-structure constant $\alpha$ is expected to be higher than in neutral atoms as a consequence of strong relativistic effects and high ionization energies~\cite{BerDzuFla10}. Such a sensitivity to $\alpha$ variation is essential to search for hypothetical oscillations and occasional jumps of $\alpha$ due to  topological defects~\cite{DerPos14} and cosmological fields, including dark matter~\cite{StaFla15,StaFla16}.

The theoretical efforts were supported by the development of experimental techniques allowing to decelerate, trap, cool, and control HCIs.
It was demonstrated that HCIs produced in an electron beam ion trap (EBIT) can be ejected, decelerated, and stopped inside of a Coulomb crystal of laser-cooled Be$^+$ ions confined in a cryogenic Paul trap \cite{schmoger_deceleration_2015,schwarz_cryogenic_2012}. Sympathetic cooling allowed to decrease the temperature of HCIs to a mK regime~\cite{SchVerSch15}. The sympathetic cooling of a single Ar$^{13+}$ to the motional ground state was demonstrated in a new cryogenic Paul trap experiment at PTB~\cite{leopold_cryogenic_2019,micke_closed-cycle_2019}. In 2020, coherent laser spectroscopy of highly
charged  $^{40}\!$Ar$^{13+}$ using quantum logic was demonstrated, achieving an increase in precision of HCI frequency measurement by eight orders of magnitude \cite{MicLeoKin20}.

In this work we explore a possibility to develop optical clocks using the transitions between the ground and a low-lying excited state
of the highly-charged Cf$^{15+}$ and Cf$^{17+}$ ions. Three out of eight main Cf isotopes have a long half-life: $A=249, I=9/2$ (351 y),
$A=250, I=0$ (13.1 y), and $A=251, I=1/2$ (898 y), where $A$ is the number of nucleons and $I$ is the nuclear spin.

Both Cf$^{15+}$ and Cf$^{17+}$ ions have the $[1s^2, ..., 5d^{10}, 6s^2]$ core. The former, Cf$^{15+}$, is a Bi-like ion with three valence electrons above the core, while Cf$^{17+}$ has one valence electron above the core, allowing to consider it as a univalent element.
But as a detailed analysis shows, more correct and accurate results are obtained if we consider Cf$^{17+}$ as a trivalent ion
including both $6s$ electrons into the valence field. This is particular important for correct determination of lowest-lying even-parity
energy levels whose main configuration, according to our calculation, is ($6s\, 5f^2$), i.e., it contains unpaired $6s$ electron.

Both the Cf$^{17+}$ and Cf$^{15+}$ ions were studied previously in Refs.~\cite{BerDzuFla12,DzuSafSaf15} and found to be particularly good candidates for testing variation of the fine-structure constant. The calculation carried out in~\cite{DzuSafSaf15} identified the ground and first excited state of Cf$^{15+}$
as the states with a high sensitivity to $\alpha$ variation and convenient clock wavelength. The dimensionless sensitivity factor $|\Delta K|$ to a variation of $\alpha$ for the Cf$^{17+}$ and Cf$^{15+}$  clock pair was predicted to be 107 (see \cite{KozSafCre18}), while the largest $|\Delta K|$ factor for any of the currently operating clock pair is 7 (for E3/E2 transitions in Yb$^+$) and most are below 1.

This paper is a guide for future experimental work, providing a detailed assessment of both ions for the clock development missing so far for most of the suggested HCI clock candidates, as noted in the recent review~\cite{KozSafCre18}.
In Sections~\ref{method} and \ref{energies} we briefly describe the method of calculation and discuss the properties of the low-lying states, such as  energies, lifetimes, and transition wavelengths. In Section~\ref{Sys_eff} we explore a number of systematic effects, such as
the electric quadrupole-, micromotion-, and quadratic Zeeman shifts of the clock transitions, which affect the accuracy of optical clocks.
We also present the results of calculation of the magnetic dipole hyperfine-structure (hfs) constants of the clock states and the black-body radiation (BBR) shifts of the clock transitions. The final section contains concluding remarks.

\section{Method of calculation}
\label{method}
We consider Cf$^{15+}$ and Cf$^{17+}$ as the ions with three valence electrons above closed cores $[1s^2,...,5d^{10} 6s^2]$ and
$[1s^2,...,5d^{10}]$, respectively. We start from solution of the Dirac-Hartree-Fock (DHF) equations in the $V^{N-3}$ approximation for both ions, where $N$ is the total number of electrons. The initial self-consistency procedure was carried out for the core electrons and then
the $5f,6p,6d,7s$, and $7p$ orbitals (and also $6s$ in case of Cf$^{17+}$) were constructed in the frozen-core potential.
The remaining virtual orbitals were formed using a recurrent procedure described in~\cite{KozPorFla96}. For both ions,
the basis sets included in total 7 partial waves ($l_{\rm max} = 6$) and orbitals with principal quantum number $n$ up to 25. We included
the Breit interaction on the same footing as the Coulomb interaction at the stage of constructing the basis set.
QED corrections were also included following Ref.~\cite{ShaTupYer13,TupKozSaf16}.

We use a hybrid approach combining configuration interaction (CI) (that takes into account an interaction between valence electrons)
and a method accounting for core-valence correlations~\cite{DzuFlaKoz96,SafKozJoh09}. The wave functions and energy levels of the valence electrons were found by solving the multiparticle relativistic equation~\cite{DzuFlaKoz96},
\begin{equation}
H_{\rm eff}(E_n) \Phi_n = E_n \Phi_n,
\label{Heff}
\end{equation}
where the effective Hamiltonian is defined as
\begin{equation}
H_{\rm eff}(E) = H_{\rm FC} + \Sigma(E),
\label{Heff1}
\end{equation}
with $H_{\rm FC}$ being the Hamiltonian in the frozen-core approximation.
The energy-dependent operator $\Sigma(E)$ accounts for virtual excitations of the core electrons.
We constructed it in three ways: using (i) the second-order many-body perturbation theory (MBPT) over residual Coulomb interaction~\cite{DzuFlaKoz96},
(ii) the linearized coupled cluster single-double (LCCSD) method~\cite{Koz04,SafKozJoh09}, and
(iii) the coupled cluster single double (valence) triple (CCSDT) method.
In the last case, using the expressions for cluster amplitudes derived in~\cite{PorDer06}, we included the non-linear (NL) terms and valence triple excitations into the formalism of the CI+all-order method developed in Ref.~\cite{SafKozJoh09}. We note that the equations for the valence triples are
solved iteratively. In the following we refer to these approaches, as the CI+MBPT, CI+LCCSD, and CI+CCSDT methods.
 \begin{center}
\begin{figure}[t]
            \includegraphics[scale=0.75]{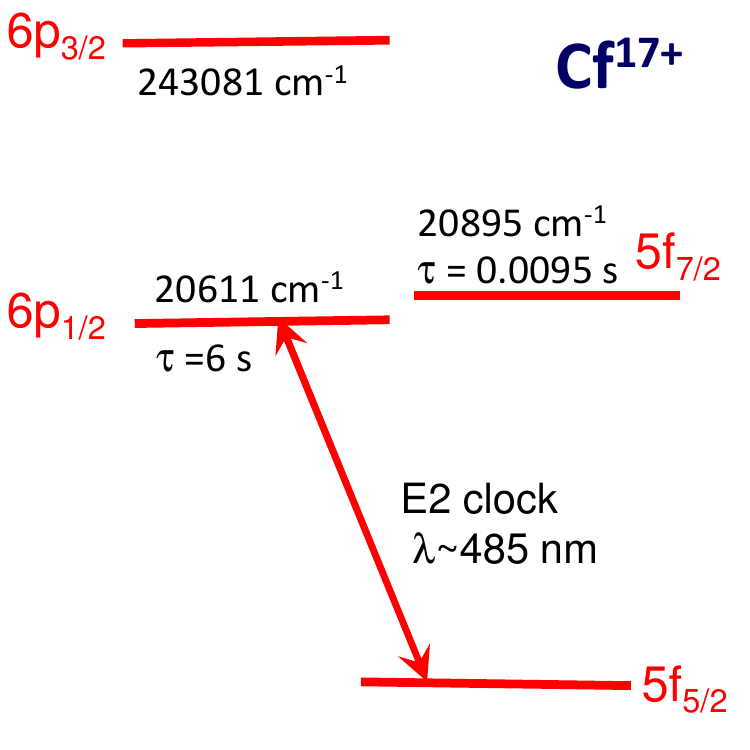}
            \caption{The level scheme for low-lying odd-parity levels of Cf$^{17+}$. }
\label{fig1}
\end{figure}
\end{center}

The sets of Cf$^{15+}$ configurations for the odd- and even-parity states were constructed by allowing single and double excitations from the $5f 6p^2$ and $5f^2 6p$ configurations and from the $6p^2 6d$, $5f 6p 6d$ and $5f^2 6d$ configurations, respectively, to $7-20s$, $7-20p$, $7-20d$, $6-19f$, and $5-13g$ shells (we designate it as $[20spd19f13g]$).
The sets of Cf$^{17+}$ configurations for the odd- and even-parity states were formed allowing single and double excitations from the
$6s^2 5f$ and $6s^2 6p$ and from the $6s 5f^2$ and $6s 5f 6p$ configurations, respectively, to $[20spd19f13g]$.
We checked for both ions that if we allowed the single and double excitations to higher lying $f$ and $g$ shells and also triple excitations
from the main configurations, the energies (counted from the ground state) changed only by few tens cm$^{-1}$. 

\begin{center}
\begin{figure}[ht]
            \includegraphics[scale=0.65]{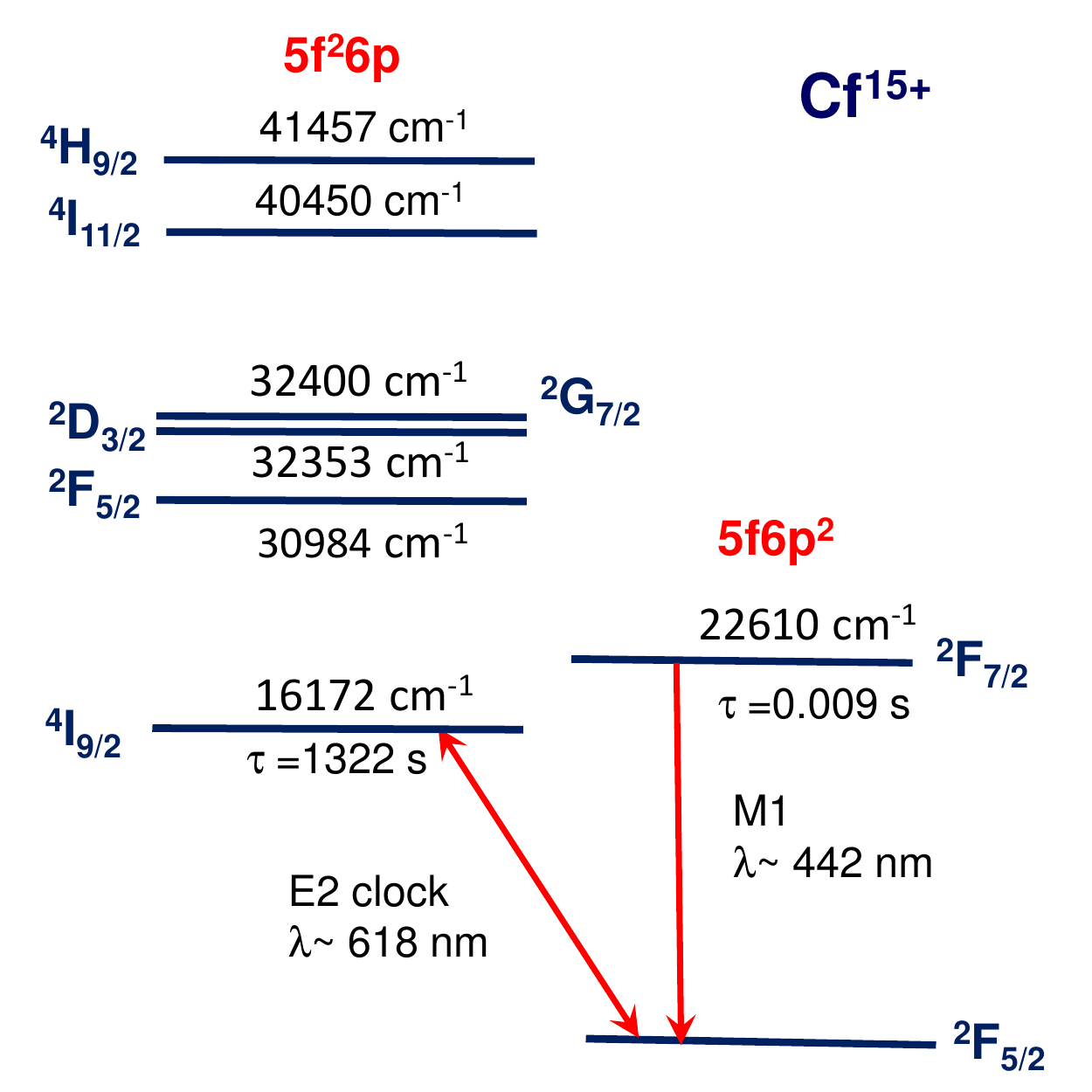}
            \caption{The level scheme for low-lying levels of  Cf$^{15+}$. }
\label{fig2}
\end{figure}
\end{center}

The level schemes for low-lying levels of Cf$^{17+}$ and Cf$^{15+}$ are given in Fig.~\ref{fig1} and
Fig.~\ref{fig2}.
\section{Energy levels}
\label{energies}
The energies of the lowest-lying states of Cf$^{15+}$ and Cf$^{17+}$ obtained in different approximations are listed in~\tref{Energ}.
The energies of the excited states (in cm$^{-1}$) are counted from the ground state. The assignments of the Cf$^{15+}$ odd levels
are from Ref.~\cite{DzuSafSaf15}. For designation of all other terms we use the main configuration and the total angular momentum $J$ of the state
as a subscript.
\begin{table*}[tp]
\caption{The energies of the excited states (in cm$^{-1}$), counted from the ground state, calculated
in the CI and CI+MBPT approximations. Contributions from higher-order (HO) correlations (difference of the CI+LCCSD and CI+MBPT calculations)
and from the NL terms and triple excitations (difference of the CI+CCSDT and CI+LCCSD calculations) and estimated contributions of
higher partial waves ($l > 6$) are given separately in columns HO, NLTr, and Extrap. The final values, given in the column labeled ``Final'',
are obtained as the sum of the CI+MBPT values and HO, NLTr, and Extrap corrections. We use the main configuration and the total angular momentum $J$ as a subscript to designate the Cf$^{15+}$ even-parity levels and the levels of Cf$^{17+}$. }
\label{Energ}%
\begin{ruledtabular}
\begin{tabular}{ccccccccccc}

&\multicolumn{1}{c}{Level} & \multicolumn{1}{c}{CI} & \multicolumn{1}{c}{CI+MBPT} & \multicolumn{1}{c}{HO}
& \multicolumn{1}{c}{NLTr} & \multicolumn{1}{c}{Extrap} &  \multicolumn{1}{c}{Final} & \multicolumn{1}{c}{Ref.~\cite{BerDzuFla12}}
& \multicolumn{1}{c}{Ref.~\cite{TupKozSaf16}} & \multicolumn{1}{c}{Ref.~\cite{DzuSafSaf15}} \\
\hline      \\[-0.6pc]
Cf$^{15+}$ & $5f 6p^2\,\,^2\!F^o_{5/2}$  &    0    &    0    &    0    &    0    &    0    &    0    &        &    0   &    0    \\[0.1pc]
           & $5f^2 6p\,\,^4\!I^o_{9/2}$  &  28930  &  10549  &   2907  &   3675  &   -959  &  16172  &        &  12898 &  12314  \\[0.1pc]
           & $5f 6p^2\,\,^2\!F^o_{7/2}$  &  22269  &  22388  &   -107  &    486  &   -158  &  22610  &        &  22018 &  21947  \\[0.1pc]
           & $5f^2 6p\,\,^2\!F^o_{5/2}$  &  43441  &  25803  &   2242  &   3741  &   -802  &  30984  &        &  27127 &  26665  \\[0.1pc]
           & $5f^2 6p\,\,^2\!D^o_{3/2}$  &  45515  &  26984  &   2483  &   3855  &   -969  &  32353  &        &        &  27750  \\[0.1pc]
           & $5f^2 6p\,\,^2\!G^o_{7/2}$  &  43552  &  28809  &   1276  &   3081  &   -765  &  32400  &        &  29214 &  28875  \\[0.1pc]
           & $5f^2 6p\,\,^4\!I^o_{11/2}$ &  51995  &  35979  &   1715  &   3717  &   -961  &  40450  &        &  37081 &  36564  \\[0.1pc]
           & $5f^2 6p\,\,^4\!H^o_{9/2}$  &  52793  &  37304  &   1522  &   3564  &   -934  &  41457  &        &  37901 &  37392  \\[0.3pc]

           & $(6p^2 6d)_{3/2}$           & 520444  & 544228  &  -3419  &  -4383  &   1089  & 537515  &        &        &         \\[0.1pc]
           & $(5f 6p 6d)_{9/2}$          & 534519  & 545581  &  -1612  &  -2445  &    249  & 541773  &        &        &         \\[0.1pc]
           & $(5f 6p 6d)_{7/2}$          & 538082  & 548797  &  -1634  &  -2152  &    235  & 545245  &        &        &         \\[0.1pc]
           & $(5f 6p 6d)_{5/2}$          & 538863  & 549387  &  -1508  &  -2156  &    216  & 545939  &        &        &         \\[0.1pc]
           & $(5f 6p 6d)_{3/2}$          & 547123  & 556562  &  -1207  &  -1907  &    190  & 553637  &        &        &         \\[0.5pc]

Cf$^{17+}$ & $(6s^2\,\,5f)^o_{5/2}$      &   0     &    0    &     0   &    0    &     0   &    0    &    0   &        &         \\[0.1pc]
           & $(6s^2\,\,6p)^o_{1/2}$      & 10104   &  22118  &  -1402  &  -1126  &  1021  &  20611  &  18686 &        &         \\[0.1pc]
           & $(6s^2\,\,5f)^o_{7/2}$      & 19682   &  22116  &  -1102  &   -152  &    33  &  20895  &  21848 &        &         \\[0.1pc]
           & $(6s^2\,\,6p)^o_{3/2}$      & 228778  & 245070  &  -1783  &  -1341  &  1136  & 243081  & 242811 &        &         \\[0.3pc]

           & $(6s\,\,5f^2)_{7/2}$        & 206421  & 202671  &   -496  &   -340  &  -945  & 200890  &        &        &         \\[0.1pc]
           & $(6s\,\,5f^2)_{9/2}$        & 211719  & 208829  &   -707  &   -415  &  -942  & 206765  &        &        &         \\[0.1pc]
           & $(6s\,\,5f^2)_{3/2}$        & 212749  & 210608  &   -860  &   -414  &  -833  & 208501  &        &        &         \\[0.1pc]
           & $(6s\,\,5f^2)_{5/2}$        & 219342  & 213728  &  -1663  &  -1524  &  -359  & 210182  &        &        &         \\[0.1pc]
           & $(6s\,5f\,6p)_{5/2}$        & 206500  & 220621  &  -1855  &  -1252  &  -463  & 217050  &        &        &
\end{tabular}
\end{ruledtabular}
\end{table*}

In the third and forth columns we present the pure CI and CI+MBPT values.
Contributions from higher-order (HO) correlations (difference of the CI+LCCSD and CI+MBPT calculations)
and from the NL terms and triple excitations (difference of the CI+CCSDT and CI+LCCSD calculations) are given separately in columns
labeled ``HO'' and ``NLTr''. Following an empiric rule obtained for Ag-like ions in Ref.~\cite{SafDzuFla14PRA1} and applied for Cd-like and Sn-like ions
in Ref.~\cite{SafDzuFla14PRA2} we estimate the contribution of the higher ($l > 6$) partial waves as the contribution of
the $l = 6$ partial wave obtained as the difference of two calculations where all intermediate sums in the all-order and
MBPT terms are restricted to $l_{\rm max} = 6$ and $l_{\rm max} = 5$. This contribution is listed in~\tref{Energ} in column labeled ``Extrap''.
The final theoretical results, listed in the ?œFinal?? column, are obtained as the sum of the CI+MBPT values and HO, NLTr, and Extrap corrections.

We find that the clock transition energies between the ground and first excited state are very sensitive to different corrections
for both ions. The CI+MBPT value differs from the CI value by more than a factor of 2 for both ions, i.e., the
contribution of the core-valence correlation corrections is as large as the CI result. An inclusion of the HO corrections, the NL terms
and valence triples in the framework of the CI+LCCSD and CI+CCSDT methods further changed the energies by several thousands of cm$^{-1}$. 

The Cf$^{15+}$ clock transition energy found at the CI+LCCSD stage is in a reasonable agreement with the results of Refs.~\cite{TupKozSaf16,DzuSafSaf15}. The quadratic NL terms and valence triples, contributing 3675 cm$^{-1}$ to the transition energy,
were not taken into account in~\cite{TupKozSaf16,DzuSafSaf15}, what explains a difference between the present result and the clock transition energy predicted in those works.
Taking into account an importance of the NL terms and valence triple excitations and also noting that the present calculation still omits
the core triples and higher-order NL terms, we estimate the uncertainty
of the clock transition energies as a half of difference between the CI+CCSDT and CI+LCCSD values. 

This conservative estimate is based on a
conclusion drawn from calculations for Na~\cite{CanDer04} and Cs~\cite{DerPor05} that the contribution from the valence triples
and NL terms is (much) larger than the contribution from core triples. Thus, the uncertainty of the clock
transition energy is $\sim\! 1800\,\, {\rm cm}^{-1}$ for Cf$^{15+}$ and $\sim\! 600\,\, {\rm cm}^{-1}$ for Cf$^{17+}$. Taking these uncertainties into account we neglect corrections to the transition energies due to effective three-particle interactions between valence electrons. These corrections were found to be at the level of 100 cm$^{-1}$ or less for the low-lying states of Cf$^{15+}$~\cite{TupKozSaf16}.

In \tref{tau} we present the wavelengths between the ground and excited states (in nm) and the excited states lifetimes (in s)
for Cf$^{15+}$ and Cf$^{17+}$ obtained in the CI+CCSDT approximation and compare with other calculations where available.
The Cf$^{15+}$ first excited state,
$5f^2 6p\,\,^4\!I^o_{9/2}$, has a rather long lifetime, 22 min. This is because it decays to the ground state
through a weak $E2$ transition.
Our predicted lifetime of the $^4\!I^o_{9/2}$ state is 5 times smaller than the value obtained in Ref.~\cite{DzuSafSaf15}, mostly
due to change in the predicted clock transition energy, since the probability of the $E2$ transition is proportional to $(\Delta E)^5$.
The lifetimes of other listed excited states are several orders of magnitude smaller. In particular,
$5f 6p^2\,\,^2\!F^o_{5/2}$ and $^2\!F^o_{7/2}$ are the fine-structure levels of the same manifold and
there is a relatively strong $M1$ $^2\!F^o_{7/2} -\,^2\!F^o_{5/2}$ transition. The same is true for the $5f^2 6p\,\,^4\!I^o_{9/2}$ and $^4\!I^o_{11/2}$ pair of levels.
\begin{table}[tp]
\caption{The wavelengths between the ground and excited states (in nm) and the excited states lifetimes (in s).}
\label{tau}%
\begin{ruledtabular}
\begin{tabular}{cccccc}
&& \multicolumn{2}{c}{This work} & \multicolumn{2}{c}{Ref.~\cite{DzuSafSaf15}} \\
&\multicolumn{1}{c}{Level} &\multicolumn{1}{c}{$\lambda$(nm)} &\multicolumn{1}{c}{$\tau$(s)}
&\multicolumn{1}{c}{$\lambda$(nm)} &\multicolumn{1}{c}{$\tau$(s)} \\
\hline      \\[-0.6pc]
Cf$^{15+}$ & $5f 6p^2\,\,^2\!F^o_{5/2}$  &   0   &        &   0   &  0    \\[0.1pc]
           & $5f^2 6p\,\,^4\!I^o_{9/2}$  &  618  & 1322   &  812  & 6900  \\[0.1pc]
           & $5f 6p^2\,\,^2\!F^o_{7/2}$  &  442  & 0.009  &  456  & 0.012 \\[0.1pc]
           & $5f^2 6p\,\,^2\!F^o_{5/2}$  &  323  & 0.18   &  375  & 0.26  \\[0.1pc]
           & $5f^2 6p\,\,^4\!I^o_{11/2}$ &  247  & 0.003  &  273  & 0.003 \\[0.3pc]

Cf$^{17+}$ & $(6s^2\,\,5f)^o_{5/2}$      &   0   &        &       &       \\[0.1pc]
           & $(6s^2\,\,6p)^o_{1/2}$      & 485  &  6.0    &       &       \\[0.1pc]
           & $(6s^2\,\,5f)^o_{7/2}$      &  479  & 0.0095 &       &       \\[0.1pc]
           & $(6s^2\,\,6p)^o_{3/2}$      &   41  & 7$\times 10^{-6}$ &       &
\end{tabular}
\end{ruledtabular}
\end{table}


For Cf$^{17+}$, the $6s^2\,6p_{1/2}$ clock excited state also decays to the ground state through the $E2$ transition. The probability of this transition
is 0.17 s$^{-1}$ leading to the lifetime of this state, $\tau \approx 6.0$ s.
We note that the probability of the $M3$ $6s^2\,6p_{1/2} -\, 6s^2\,5f_{7/2}$ transition is negligible.
\section{Systematic effects}
\label{Sys_eff}
In this section we consider a number of systematic effects relevant to the clock $5f^2 6p\,\, ^4\!I^o_{9/2} -\,5f 6p^2\,\, ^2\!F^o_{5/2}$ and
$6s^2\,6p_{1/2} -\, 6s^2\,5f_{5/2}$ transitions in Cf$^{15+}$ and Cf$^{17+}$, respectively. We use wave functions
obtained in the CI+CCSDT approximation in all subsequent calculations for both ions. We also simplify notation for the Cf$^{17+}$ clock states as $6s^2\,5f_{5/2} \equiv 5f_{5/2}$
and $6s^2\,6p_{1/2} \equiv 6p_{1/2}$. In calculating matrix elements (MEs) of different operators the random phase approximation (RPA) corrections
were included.
\subsection{Electric quadrupole shift}
The Hamiltonian, $H_Q$, describing the interaction of the external electric-field gradient with the quadrupole
moment of an atomic state $|\gamma JIFM \rangle$ (where $J$ is the total angular momentum of the electrons, $I$ is the nuclear spin,
{\bf F} = {\bf J} + {\bf I}, M is the projection of {\bf F}, and $\gamma$ encapsulates all other electronic quantum numbers) is given by
\begin{equation}
H_Q = \sum_{q=-2}^2 (-1)^q \nabla \mathcal{E}^{(2)}_q Q_{-q} ,
\label{HQ}
\end{equation}
where the $q=0$ component of $\nabla \mathcal{E}^{(2)}$ can be written as~\cite{Ram56,Ita00}:
\begin{equation}
\nabla \mathcal{E}^{(2)}_0 = -\frac{1}{2}\,\frac{\partial {\mathcal E}_z}{\partial z} .
\end{equation}
Coupling of this field gradient to the quadrupole moment of the atomic state leads to the energy shift:
\begin{equation}
\Delta E = -\frac{1}{2}\, \langle Q_0 \rangle \,\frac{\partial {\mathcal E}_z}{\partial z} ,
\end{equation}
where $\langle Q_0 \rangle \equiv \langle \gamma JIFM |Q_0| \gamma JIFM \rangle$.

The fractional electric quadrupole shift of the clock transition is then
\begin{equation}
\frac{\Delta \nu}{\nu_{\rm clock}} = -\frac{1}{2 h \nu_{\rm clock}}\,
\Delta \langle Q_0 \rangle \, \frac{\partial {\mathcal E}_z}{\partial z} ,
\label{del_om}
\end{equation}
where $\nu_{\rm clock}$ is the clock transition frequency, $h$ is the Planck constant, and $\Delta \langle Q_0 \rangle$ is the difference of the expectation values of $Q_0$ for the upper and lower clock states.

The ME $\langle \gamma JIFM |Q_0| \gamma JIFM \rangle$ can be written as
\begin{eqnarray}
&&\langle \gamma JIFM |Q_0| \gamma JIFM \rangle = (-1)^{I+J+F} \nonumber \\
&\times&[3M^2 - F(F+1)] \sqrt{\frac{2F+1}{(2F+3)(F+1)F(2F-1)}} \nonumber \\
&\times& \left\{
\begin{array}{ccc}
J & 2 & J \\
F & I & F
\end{array}
\right\}
\langle \gamma J ||Q|| \gamma J \rangle ,
\label{Q0}
\end{eqnarray}
where $\langle \gamma J ||Q|| \gamma J \rangle$ is the reduced ME of the electric quadrupole operator.

Our calculation gives
\begin{eqnarray}
\langle ^2\!F^o_{5/2} ||Q|| ^2\!F^o_{5/2} \rangle &\approx& 0.31\, |e|\, a_0^2 , \nonumber \\
\langle ^4\!I^o_{9/2} ||Q|| ^4\!I^o_{9/2} \rangle &\approx& 0.53\, |e|\, a_0^2 ,
\label{Q_ME}
\end{eqnarray}
for the  ground and first excited states of Cf$^{15+}$, where $e$ is the electron charge and $a_0$ is the Bohr radius.

Using these MEs and the expression for the quadrupole moment $\Theta$ of an atomic state $|\gamma J \rangle$ given by
\begin{eqnarray}
\Theta &=& 2\, \langle \gamma J, M_J=J |Q_0| \gamma J, M_J=J \rangle \nonumber \\
       &=& 2\, \sqrt{\frac{J(2J-1)}{(2J+3)(J+1)(2J+1)}} \langle \gamma J ||Q|| \gamma J \rangle
\end{eqnarray}
we can find the quadrupole moments of the clock states to be
\begin{eqnarray}
\Theta (^2\!F^o_{5/2}) &\approx& 0.15\, |e|\, a_0^2 , \nonumber \\
\Theta (^4\!I^o_{9/2}) &\approx& 0.25\, |e|\, a_0^2 .
\end{eqnarray}

As follows from \eref{Q0}, the quadrupole shift turns to zero when $3M^2 = F(F+1)$.
For both fermionic 249 and 251 isotopes of Cf with $I=9/2$ and $I=1/2$, there are sublevels of the ground state with $F=3, M=\pm 2$ for which the quadrupole shift disappears.
For the 249 isotope, the total angular momentum $F$ of the upper clock state ranges from 0 to 9.
If we also choose $F=3, M=\pm 2$ for this state, the clock transition is not affected by the quadrupole shift. Averaging over
the $M=\pm 2$ transitions furthermore eliminates the linear Zeeman shift.
For the 251 isotope, the upper state total angular momentum $F$ can be equal to 4 or 5.
Averaging over all pairs of $\pm|M|$ in this excited state will make the difference $3M^2 - F(F+1)$ vanish to suppress the electric quadrupole shift~\cite{dube_electric_2005}.

In general, as follows from~\eref{Q0},
\begin{equation}
\sum_M \langle \gamma JIFM |Q_0| \gamma JIFM \rangle =0,
\end{equation}
and the same is true for the $H_Q$ operator given by \eref{HQ}~\cite{Ita00}.
Thus, the quadrupole shift vanishes when averaged over all $M$.
This technique has been employed in singly-charged frequency standards \cite{margolis_hertz-level_2004, chwalla_absolute_2009, MadDubZho12}
to suppress the uncertainty in this shift by up to four orders of magnitude \cite{DubMadZho13}.

To get an upper limit for the quadrupole shift we put $M=0$ in \eref{Q0} and chose such values of $F$ for the upper and lower clock states
to maximize $|\Delta \langle Q_0 \rangle|$. It gives us $|\Delta \langle Q_0 \rangle| \sim 0.1\, |e|\, a_0^2$.
Substituting it into \eref{del_om} and using for an estimate
$\partial {\mathcal E}_z/{\partial z} \approx 1\,\, {\rm kV}/{\rm cm}^2 \approx 1.029 \times 10^{-15}\,\, {\rm a.u.}$, we obtain
for the quadrupole shift:
\begin{equation}
\frac{\Delta \nu}{\nu_{\rm clock}} \simeq 7 \times 10^{-16} .
\end{equation}
Even in this (worst) case,  a 3-4 order of magnitude suppression will make the electric quadrupole shift well below
 $10^{-18}$.

For Cf$^{17+}$, the quadrupole moment of the upper clock state $6p_{1/2}$ is equal to 0. For the ground $5f_{5/2}$ state we obtain
\begin{eqnarray}
\langle 5f_{5/2} ||Q|| 5f_{5/2} \rangle &\approx& 0.80\, |e|\, a_0^2 , \nonumber \\
\Theta (5f_{5/2}) &\approx& 0.39\, |e|\, a_0^2 .
\label{Q_ME1}
\end{eqnarray}

For both 249 and 251 isotopes there is the sublevel of the ground state with $F=3, M=\pm 2$
for which the quadrupole shift vanishes. As a result, it vanishes also for the clock transition.

We can compare these results with those obtained for Sr$^{+}$ where the suppression technique discussed above was applied.
Using the recent measurement of the electric quadrupole moment of the Sr$^{+}$ $4d_{5/2}$ clock state~\cite{ShaAkeOze16}
and noting that our definition of the quadrupole moment differs by factor of 2 from that used in Ref.~\cite{ShaAkeOze16},
we obtain $|\langle 4d_{5/2} ||Q|| 4d_{5/2} \rangle| \approx 10.7\, |e|\, a_0^2$.
This value is more than an order of magnitude larger than the respective MEs for Cf$^{15+}$ and Cf$^{17+}$
given by Eqs.~(\ref{Q_ME}) and (\ref{Q_ME1}).

\subsection{Black-body radiation shift}
A BBR shift of the clock energy levels is due to an interaction of thermal photons
with the atom. The fractional shift of the clock transition is given by
\begin{eqnarray}
\frac{\Delta \nu_{\rm BBR}}{\nu_{\rm clock}} &\approx&
-\frac{\pi^2}{15\, c^3 \hbar^4} \frac{\Delta \alpha}{\nu_{\rm clock}}\,(k_B T)^4  \nonumber \\
&\equiv& \beta_{\rm BBR} \left(\frac{T}{300\, K}\right)^4 ,
\label{BBR}
\end{eqnarray}
where $\Delta \alpha \equiv \alpha(^4\!I^o_{9/2}) - \alpha(^2\!F^o_{5/2})$ for Cf$^{15+}$ and
$\Delta \alpha \equiv \alpha(6p_{1/2}) - \alpha(5f_{5/2})$ for Cf$^{17+}$ are the differential scalar static polarizabilities,
$c$ is the speed of light, $k_B$ is the Boltzmann constant, and $T$ is the BBR temperature.

We can present the scalar polarizability $\alpha$ as a sum of the valence polarizability, $\alpha_v$, ionic-core polarizability $\alpha_c$,
and a small term $\alpha_{vc}$ that modifies ionic-core polarizability due to the presence of valence electrons:
\begin{eqnarray}
\alpha = \alpha_v + \alpha_c + \alpha_{vc} .
\end{eqnarray}
The valence part of the scalar static polarizability of a state $|0\rangle$ with the energy $E_0$ and total angular momentum $J_0$
is determined as
\begin{equation}\label{genpol}
\alpha_0^v = \frac{2}{3(2J_0+1)}\sum_n \frac{|\langle 0\| D\| n\rangle|^2}{E_n-E_0} ,
\end{equation}
where $\bf D$ is the electric-dipole operator.
Instead of direct summation over all intermediate states we solve the inhomogeneous equation in the valence space~\cite{KozPor99a}:
\begin{equation}
(E_0 - H_{\textrm{eff}})|\psi \rangle = D_z |0\rangle
\end{equation}
and then use $|\psi \rangle$ to find $\alpha_0^v$.
The core and $vc$ terms are evaluated in the single-particle approximation including RPA~\cite{SafJohDer99};
$\alpha_{vc}$ are calculated by adding $vc$ contributions from individual electrons.
Thus, for Cf$^{15+}$,
$\alpha_{vc} (^2\!F^o_{5/2}) = \alpha_{vc}(5f_{5/2}) + 2 \alpha_{vc}(6p_{1/2})$ and
$\alpha_{vc} (^4\!I^o_{9/2}) = 2 \alpha_{vc}(5f_{5/2}) + \alpha_{vc}(6p_{1/2})$.

The results of calculation of the scalar static polarizabilities and the parameters $\beta_{\rm BBR}$ are given in \tref{polariz}.
Only the valence polarizabilities $\alpha_v$ were found in Ref.~\cite{DzuSafSaf15}; these results are in a reasonable agreement
with our values for $\alpha_v$. We would like to note an enhanced role of the $vc$ terms for Cf$^{15+}$.
While the core contribution cancels in the differential polarizability, the $vc$ term does not. It nearly cancels the valence polarizability and significantly affects the result.

\begin{table}[tp]
\caption{Contributions $\alpha_v$, $\alpha_c$, and $\alpha_{vc}$ to the scalar static polarizabilities of the clock states,
the differential polarizabilities $\Delta \alpha$, and the parameter $\beta_{\rm BBR}$, determined in the text, are presented.
$\alpha = \alpha_v + \alpha_c + \alpha_{vc}$.}
\label{polariz}%
\begin{ruledtabular}
\begin{tabular}{ccccc}
& \multicolumn{1}{c}{State} & \multicolumn{1}{c}{(in $a_0^3$)} & \multicolumn{1}{c}{This work} & \multicolumn{1}{c}{Ref.~\cite{DzuSafSaf15}} \\
\hline      \\[-0.6pc]
Cf$^{15+}$ & $^2\!F^o_{5/2}$ & $\alpha_v$    &  0.323  &    0.317\\[0.1pc]
           &                 & $\alpha_c$    &  0.948  &         \\[0.1pc]
           &                 & $\alpha_{vc}$ & -0.381  &         \\[0.1pc]
           &                 & $\alpha$      &  0.890  &  		 \\[0.3pc]

           & $^4\!I^o_{9/2}$ & $\alpha_v$    &  0.245  &  0.183  \\[0.1pc]
           &                 & $\alpha_c$    &  0.948  &         \\[0.1pc]
           &                 & $\alpha_{vc}$ & -0.207  &         \\[0.1pc]
           &                 & $\alpha$      &  0.986  &         \\[0.3pc]

           &                 &$\Delta \alpha$&  0.096  & -0.134  \\[0.1pc]
           &                 &$\beta_{\rm BBR}$& $-1.7 \times 10^{-18}$ & $2.9 \times 10^{-18}$ \\
\hline      \\[-0.6pc]
Cf$^{17+}$ & $5f_{5/2}$      & $\alpha_v$    &  0.645  &         \\[0.1pc]
           &                 & $\alpha_c$    &  0.344  &         \\[0.1pc]
           &                 & $\alpha_{vc}$ & -0.028  &         \\[0.1pc]
           &                 & $\alpha$      &  0.961  &         \\[0.3pc]

           & $6p_{1/2}$      & $\alpha_v$    &  0.595  &         \\[0.1pc]
           &                 & $\alpha_c$    &  0.344  &         \\[0.1pc]
           &                 & $\alpha_{vc}$ & -0.020  &         \\[0.1pc]
           &                 & $\alpha$      &  0.919  &         \\[0.3pc]

           &                 &$\Delta \alpha$& -0.042  &         \\[0.1pc]
           &                 &$\beta_{\rm BBR}$& $5.9 \times 10^{-19}$ &
\end{tabular}
\end{ruledtabular}
\end{table}

The differential polarizabilities, $\Delta \alpha$, are very small for both ions leading to small values of the static BBR shifts (we neglect dynamic corrections to them).
We note that for both ions, the scalar static polarizabilities of the clock states are close in magnitude and by an order of magnitude (in absolute value)
larger than $\Delta \alpha$. As a result the uncertainty of the differential polarizabilities is large. For instance for Cf$^{17+}$,
if $\alpha(6p_{1/2})$ is increased by 1\% while $\alpha(5f_{5/2})$ is reduced by 1\%, $\Delta \alpha$ will change by a factor of 2.
Thus, one should consider the values of the differential polarizabilities as estimates.

The BBR shifts of the Cf$^{15+}$ and Cf$^{17+}$ clock transitions  of the order of $10^{-18}$ even at $T=300\,{\rm K}$. Since the highly-charged ion trap is operated at cryogenic temperature near
$T = 4\,{\rm K}$~\cite{KozSafCre18} the BBR shifts for both ions will be suppressed by more than 7 orders of magnitude,
even compared to small room temperature values, making them  negligible.
\subsection{Micromotion shift}
A micromotion driven by the rf-trapping field leads to ac Stark and second-order Doppler shifts.
As it was shown in~\cite{DubMadZho13}, if $\Delta \alpha$ for the clock transition is negative, there is a ``magic'' trap drive frequency
$\Omega$ given by
\begin{equation}
\Omega = \frac{|e|}{M_i c}\sqrt{-\frac{h \nu_{\rm clock}}{\Delta \alpha}}
\label{Omega}
\end{equation}
($M_i$ is the ion mass) at which the micromotion shift vanishes.
Substituting $M_i \approx A\, m_p$ (where $m_p$ is the proton mass and we use for an estimate $A=251$) and the differential polarizability of Cf$^{17+}$,
$\Delta \alpha = -0.042 \, a_0^3$, to \eref{Omega}, we obtain $\Omega \approx 2\,\pi \times 155\,\, {\rm MHz}$.

For Cf$^{15+}$ we obtained positive value of $\Delta \alpha$ and a ``magic'' trap drive frequency does not exist.
But in this case compensation voltages, allowing to direct the ion back to a position where radio-frequency field vanishes,
can be applied~\cite{BerMilBer98,KelParBur15}.
If these voltages are well controlled the excess micromotion does not pose a limitation to optical frequency standards~\cite{KozSafCre18}.
\subsection{Hyperfine interaction}
We also calculated the magnetic-dipole hfs constants $A$ for
the clock states of the Cf$^{15+}$ and Cf$^{17+}$ ions.

The nuclear magnetic moment, $\mu_I$, is unknown for $^{251}$Cf. For the 249 isotope the results obtained for $\mu_I$ are somewhat contradictory.
The absolute value, $|\mu_I| = 0.28(6)\,\mu_N$ (where $\mu_N$ is the nuclear magneton), was experimentally found in~\cite{EdeKar75} while the theoretical calculation carried out in that work gave $\mu_I = -0.49\,\mu_N$.
For this reason, we present our results in the form $A/g_I$, keeping the nuclear $g$ factor, $g_I \equiv \mu_I/(I\,\mu_N)$, as a multiplier. The
values of $A/g_I$, which are approximately the same for both 249 and 251 isotopes, are listed in~\tref{hfs}. We estimate the accuracy of
these values at the level of 20-30\%.
\begin{table}[tp]
\caption{The values of $A/g_I$ (in MHz) for the clock states of Cf$^{15+}$ and Cf$^{17+}$.}
\label{hfs}%
\begin{ruledtabular}
\begin{tabular}{ccc}
& \multicolumn{1}{c}{State} & \multicolumn{1}{c}{$A/g_I$}  \\
\hline      \\[-0.6pc]
Cf$^{15+}$ & $^2\!F^o_{5/2}$ &   4200 \\[0.1pc]
           & $^4\!I^o_{9/2}$ &  21000 \\[0.3pc]

Cf$^{17+}$ & $5f_{5/2}$      &   1900 \\[0.1pc]
           & $6p_{1/2}$      & 195000
\end{tabular}
\end{ruledtabular}
\end{table}
\subsection{Zeeman shift}
In the presence of an external magnetic field $\bf B$ atomic energy levels (and transition frequencies) experience
the linear and quadratic Zeeman shifts.
The former scales linearly with the magnetic quantum number $M$. It equals 0 at $M=0$ and can be suppressed in other
cases if the frequency is averaged over two or more transitions with linear Zeeman shifts equal in absolute value but having the
opposite signs \cite{bernard_laser_1998}.

To determine the quadratic Zeeman shift in the case of a weak magnetic field, we have to consider both  hyperfine and Zeeman interactions:
\begin{equation}
H = H_{\rm hfs} - {\boldsymbol \mu}_{\rm at} {\bf B}
\end{equation}
with ${\boldsymbol \mu}_{\rm at} = -\mu_B g_J {\bf J} - \mu_N g_I {\bf I}$. Here,
$\mu_B$ is the Bohr magneton and $g_J$ is the electron $g$ factor, given in the nonrelativistic approximation by the formula
\begin{equation}
g_J = \frac{3}{2} + \frac{S(S+1) - L(L+1)}{2J(J+1)} .
\label{g_J}
\end{equation}

Below, we estimate this effect for $^{251}$Cf, that has the nuclear spin $I=1/2$. In this case,
\begin{equation}
H_{\rm hfs} = h A\, {\bf I}{\bf J} ,
\end{equation}
where $A$ is the magnetic dipole hyperfine structure constant (in Hz).

If $I=1/2$, the total angular momentum $F=J \pm 1/2$. For the case of $J=1/2,\, F=I \pm 1/2$, the resulting energy shift was obtained by Breit and Rabi in Ref.~\cite{BreRab31}.
Following the approach of~\cite{BreRab31}, we obtain for the energy shift
\begin{eqnarray}
&&\Delta E_{F=J \pm 1/2} = -\frac{h \Delta W}{2(2J+1)} + \mu_B\, g_J\, m_F B \nonumber \\
&\pm& \frac{1}{2} \sqrt{(h \Delta W)^2 + \frac{2 m_F\, h \Delta W\, y}{J+1/2} + y^2},
\label{DelE_F}
\end{eqnarray}
where
\begin{equation*}
y \equiv (\mu_N g_I - \mu_B g_J)B
\end{equation*}
and $\Delta W \equiv A (J + 1/2)$ is the splitting (in Hz) between two hyperfine sublevels in the absence of the magnetic field.

If the magnetic field is weak, $B \sim 10^{-5}\,{\rm T}$, then $|y| \ll h \Delta W$. It
 follows from \eref{DelE_F}, that the contribution quadratic in $B$ to $\Delta E_{F=J \pm 1/2}$ (we designate it as $\Delta E_{F=J \pm 1/2}^{(2)}$) is proportional to $y^2$ and is given by
\begin{eqnarray*}
\Delta E_{F=J \pm 1/2}^{(2)} = \pm\, \frac{y^2}{4\,h \Delta W} \approx \pm\, \frac{1}{2(2J+1)}\,\frac{(\mu_B g_J)^2}{h A}\,B^2 .
\label{x2}
\end{eqnarray*}

For the Cf$^{15+}$ clock $^4\!I^o_{9/2} -\, ^2\!F^o_{5/2}$ transition, we have
$g_J(^2\!F^o_{5/2}) = 6/7$ and $g_J(^4\!I^o_{9/2}) = 8/11$. Using the values of $A/g_I$ given in~\tref{hfs} for the clock states we obtain after simple transformations the frequency shift for the $^4\!I^o_{9/2}(F=5) -\, ^2\!F^o_{5/2}(F=3)$ transition,
\begin{eqnarray*}
|\Delta \nu| = \frac{|\Delta E^{(2)}(^4\!I^o_{9/2}) - \Delta E^{(2)}(^2\!F^o_{5/2})|}{h}
         \approx 2.6\, g_I \frac{{\rm kHz}}{({\rm mT})^2} B^2 .
\end{eqnarray*}

Given $B = 10\,{\rm \mu T}$, putting $g_I=1$, and using $\nu_{\rm clock} \approx 4.8 \times 10^{14}$ Hz we arrive at the estimate for
the Cf$^{15+}$ fractional clock shift:
\begin{equation}
\frac{|\Delta \nu|}{\nu_{\rm clock}} \approx 5 \times 10^{-16}.
\end{equation}
As follows from \eref{DelE_F}, for the $^4\!I^o_{9/2}(F=4) -\, ^2\!F^o_{5/2}(F=2)$ transition we will get exactly the same frequency shift $\Delta \nu$
as for the $^4\!I^o_{9/2}(F=5) -\, ^2\!F^o_{5/2}(F=3)$ transition in absolute value but with the opposite sign. Thus, an averaging of the quadratic Zeeman shifts over these two transitions will lead to complete cancellation of this effect.

Similarly, the Cf$^{17+}$ clock $6p_{1/2} -\, 5f_{5/2}$ transition frequency shift  is
\begin{eqnarray}
|\Delta \nu| &=& \frac{1}{h} \left|\Delta E^{(2)}(6p_{1/2}) - \Delta E^{(2)}(5f_{5/2}) \right| .
\end{eqnarray}

Taking into account that $A(6p_{1/2})$ is two orders of magnitude larger than $A(5f_{5/2})$ we can neglect $\Delta E^{(2)}(6p_{1/2})$
compared to $\Delta E^{(2)}(5f_{5/2})$, arriving at
\begin{eqnarray}
|\Delta \nu| \approx \frac{\left| \Delta E^{(2)}(5f_{5/2})\right|}{h} \approx 6.3\, \frac{{\rm kHz}}{({\rm mT})^2}\, g_I B^2 .
\label{Delnu}
\end{eqnarray}

Substituting $g_I=1$ and $B= 10\,\,{\rm \mu T}$ to \eref{Delnu} and using $\nu_{\rm clock} \approx 6.1 \times 10^{14}\,\, {\rm Hz}$,
we obtain for Cf$^{17+}$,
\begin{eqnarray}
\frac{|\Delta \nu|}{\nu_{\rm clock}} \approx 1.0 \times 10^{-15}.
\end{eqnarray}

At a small magnetic field of $\sim\!10\,\,{\rm \mu T}$, the fractional clock shift is with $\sim\!10^{-15}$ non-negligible for both ions. We expect that at these low fields and provided sufficient shielding, the magnetic field drifts can be reduced to a level of $<10\,\,{\rm pT}$ over time scales of several minutes. This results in relative frequency shifts of the clock transition from the linear Zeeman effect below $10^{-17}$, which can be averaged to zero by probing pairs of $\pm |m_F|$ states \cite{bernard_laser_1998}. The change in the quadratic Zeeman effect is negligible at the $10^{-20}$ level.

To determine the quadratic shift precisely, the magnetic field needs to be known with a high accuracy. The difference of frequencies
of the $|F,m_F\rangle -\, |F',m_F'\rangle$ and $|F,-m_F\rangle -\, |F',-m_F'\rangle$ hyperfine transitions will provide an accurate measurement of the $B$ field and its potential fluctuation. However, in all cases a precise measurement of the nuclear magnetic moments is required to cancel the shift, which will require a measurement of the hyperfine structure and improving the accuracy of the $A/g_I$ theoretical calculations. Alternatively, the $g$-factors can be determined using a co-trapped logic ion as a reference \cite{rosenband_observation_2007}, such as Be$^+$ with a well-known $g$-factor at the ppm level \cite{wineland_laser-fluorescence_1983}. The logic ion can also serve directly as a probe for the magnetic field during clock operation.

As we discussed above, we can eliminate the electric quadrupole shift by averaging transitions involving different Zeeman components .
The same approach can be applied, when $F=I \pm 1/2$ to eliminate the quadratic Zeeman shift.
This method works also for cancellation of the linear and quadratic Zeeman shifts in more general cases~\cite{KozSafCre18,DubMadZho13}.
\section{Conclusion}
\label{}
To conclude, we have carried out a systematic study of  the Cf$^{15+}$ and Cf$^{17+}$  properties needed for the development  of optical clocks with these ions using the hybrid approach that combines the CI and coupled cluster methods. We analysed a number of systematic effects (such as the electric quadrupole-, micromotion-, and quadratic Zeeman shifts of
the clock transitions) that affect the accuracy and stability of the optical clocks. We also calculated the hfs magnetic dipole constants of the clock states and the BBR shifts of the clock transitions. Based on our calculation and experimental progress in cooling and trapping HCIs~\cite{KozSafCre18, MicLeoKin20} we conclude that both the Cf$^{15+}$ and Cf$^{17+}$ ions are good candidates for optical clock. It was demonstrated earlier that such clocks would have very high sensitivity to a variation of the fine-structure constant \cite{BerDzuFla12,DzuSafSaf15}.

\begin{acknowledgments}
This work was supported in part by U.S. Office of Naval Research, award number N00014-17-1-2252.
S.G.P., A.I.B., M.G.K., and I.I.T. acknowledge support by the Russian
Science Foundation under Grant No.~19-12-00157. P.O.S acknowledges support from the Max-Planck?"Riken?"PTB?"Center for Time, Constants and Fundamental 
Symmetries and the Deutsche Forschungsgemeinschaft (DFG, German Research Foundation) 
through SCHM2678/5-1, and Germany?™s Excellence Strategy ?? EXC-2123/1 QuantumFrontiers 390837967.
\end{acknowledgments}

%


\begin{thebibliography}{45}%
\makeatletter
\providecommand \@ifxundefined [1]{%
 \@ifx{#1\undefined}
}%
\providecommand \@ifnum [1]{%
 \ifnum #1\expandafter \@firstoftwo
 \else \expandafter \@secondoftwo
 \fi
}%
\providecommand \@ifx [1]{%
 \ifx #1\expandafter \@firstoftwo
 \else \expandafter \@secondoftwo
 \fi
}%
\providecommand \natexlab [1]{#1}%
\providecommand \enquote  [1]{``#1''}%
\providecommand \bibnamefont  [1]{#1}%
\providecommand \bibfnamefont [1]{#1}%
\providecommand \citenamefont [1]{#1}%
\providecommand \href@noop [0]{\@secondoftwo}%
\providecommand \href [0]{\begingroup \@sanitize@url \@href}%
\providecommand \@href[1]{\@@startlink{#1}\@@href}%
\providecommand \@@href[1]{\endgroup#1\@@endlink}%
\providecommand \@sanitize@url [0]{\catcode `\\12\catcode `\$12\catcode
  `\&12\catcode `\#12\catcode `\^12\catcode `\_12\catcode `\%12\relax}%
\providecommand \@@startlink[1]{}%
\providecommand \@@endlink[0]{}%
\providecommand \url  [0]{\begingroup\@sanitize@url \@url }%
\providecommand \@url [1]{\endgroup\@href {#1}{\urlprefix }}%
\providecommand \urlprefix  [0]{URL }%
\providecommand \Eprint [0]{\href }%
\providecommand \doibase [0]{http://dx.doi.org/}%
\providecommand \selectlanguage [0]{\@gobble}%
\providecommand \bibinfo  [0]{\@secondoftwo}%
\providecommand \bibfield  [0]{\@secondoftwo}%
\providecommand \translation [1]{[#1]}%
\providecommand \BibitemOpen [0]{}%
\providecommand \bibitemStop [0]{}%
\providecommand \bibitemNoStop [0]{.\EOS\space}%
\providecommand \EOS [0]{\spacefactor3000\relax}%
\providecommand \BibitemShut  [1]{\csname bibitem#1\endcsname}%
\let\auto@bib@innerbib\@empty
\bibitem [{\citenamefont {Schm{\"o}ger}\ \emph {et~al.}(2015)\citenamefont
  {Schm{\"o}ger}, \citenamefont {Versolato}, \citenamefont {Schwarz},
  \citenamefont {Kohnen}, \citenamefont {Windberger}, \citenamefont {Piest},
  \citenamefont {Feuchtenbeiner}, \citenamefont {Pedregosa-Gutierrez},
  \citenamefont {Leopold}, \citenamefont {Micke}, \citenamefont {Hansen},
  \citenamefont {Baumann}, \citenamefont {Drewsen}, \citenamefont {Ullrich},
  \citenamefont {Schmidt},\ and\ \citenamefont {{Crespo
  L{\'o}pez-Urrutia}}}]{SchVerSch15}%
  \BibitemOpen
  \bibfield  {author} {\bibinfo {author} {\bibfnamefont {L.}~\bibnamefont
  {Schm{\"o}ger}}, \bibinfo {author} {\bibfnamefont {O.~O.}\ \bibnamefont
  {Versolato}}, \bibinfo {author} {\bibfnamefont {M.}~\bibnamefont {Schwarz}},
  \bibinfo {author} {\bibfnamefont {M.}~\bibnamefont {Kohnen}}, \bibinfo
  {author} {\bibfnamefont {A.}~\bibnamefont {Windberger}}, \bibinfo {author}
  {\bibfnamefont {B.}~\bibnamefont {Piest}}, \bibinfo {author} {\bibfnamefont
  {S.}~\bibnamefont {Feuchtenbeiner}}, \bibinfo {author} {\bibfnamefont
  {J.}~\bibnamefont {Pedregosa-Gutierrez}}, \bibinfo {author} {\bibfnamefont
  {T.}~\bibnamefont {Leopold}}, \bibinfo {author} {\bibfnamefont
  {P.}~\bibnamefont {Micke}}, \bibinfo {author} {\bibfnamefont {A.~K.}\
  \bibnamefont {Hansen}}, \bibinfo {author} {\bibfnamefont {T.~M.}\
  \bibnamefont {Baumann}}, \bibinfo {author} {\bibfnamefont {M.}~\bibnamefont
  {Drewsen}}, \bibinfo {author} {\bibfnamefont {J.}~\bibnamefont {Ullrich}},
  \bibinfo {author} {\bibfnamefont {P.~O.}\ \bibnamefont {Schmidt}}, \ and\
  \bibinfo {author} {\bibfnamefont {J.~R.}\ \bibnamefont {{Crespo
  L{\'o}pez-Urrutia}}},\ }\href@noop {} {\bibfield  {journal} {\bibinfo
  {journal} {Science}\ }\textbf {\bibinfo {volume} {347}},\ \bibinfo {pages}
  {1233} (\bibinfo {year} {2015})}\BibitemShut {NoStop}%
\bibitem [{\citenamefont {Kozlov}\ \emph {et~al.}(2018)\citenamefont {Kozlov},
  \citenamefont {Safronova}, \citenamefont {{Crespo L{\'o}pez-Urrutia}},\ and\
  \citenamefont {Schmidt}}]{KozSafCre18}%
  \BibitemOpen
  \bibfield  {author} {\bibinfo {author} {\bibfnamefont {M.~G.}\ \bibnamefont
  {Kozlov}}, \bibinfo {author} {\bibfnamefont {M.~S.}\ \bibnamefont
  {Safronova}}, \bibinfo {author} {\bibfnamefont {J.~R.}\ \bibnamefont {{Crespo
  L{\'o}pez-Urrutia}}}, \ and\ \bibinfo {author} {\bibfnamefont {P.~O.}\
  \bibnamefont {Schmidt}},\ }\href@noop {} {\bibfield  {journal} {\bibinfo
  {journal} {Rev. Mod. Phys.}\ }\textbf {\bibinfo {volume} {90}},\ \bibinfo
  {pages} {045005} (\bibinfo {year} {2018})}\BibitemShut {NoStop}%
\bibitem [{\citenamefont {{Micke}}\ \emph {et~al.}(2020)\citenamefont
  {{Micke}}, \citenamefont {{Leopold}}, \citenamefont {{King}}, \citenamefont
  {{Benkler}}, \citenamefont {{Spie{\ss}}}, \citenamefont {{Schm{\"o}ger}},
  \citenamefont {{Schwarz}}, \citenamefont {{Crespo L{\'o}pez-Urrutia}},\ and\
  \citenamefont {{Schmidt}}}]{MicLeoKin20}%
  \BibitemOpen
  \bibfield  {author} {\bibinfo {author} {\bibfnamefont {P.}~\bibnamefont
  {{Micke}}}, \bibinfo {author} {\bibfnamefont {T.}~\bibnamefont {{Leopold}}},
  \bibinfo {author} {\bibfnamefont {S.~A.}\ \bibnamefont {{King}}}, \bibinfo
  {author} {\bibfnamefont {E.}~\bibnamefont {{Benkler}}}, \bibinfo {author}
  {\bibfnamefont {L.~J.}\ \bibnamefont {{Spie{\ss}}}}, \bibinfo {author}
  {\bibfnamefont {L.}~\bibnamefont {{Schm{\"o}ger}}}, \bibinfo {author}
  {\bibfnamefont {M.}~\bibnamefont {{Schwarz}}}, \bibinfo {author}
  {\bibfnamefont {J.~R.}\ \bibnamefont {{Crespo L{\'o}pez-Urrutia}}}, \ and\
  \bibinfo {author} {\bibfnamefont {P.~O.}\ \bibnamefont {{Schmidt}}},\ }\href
  {\doibase 10.1038/s41586-020-1959-8} {\bibfield  {journal} {\bibinfo
  {journal} {\nat}\ }\textbf {\bibinfo {volume} {578}},\ \bibinfo {pages} {60}
  (\bibinfo {year} {2020})}\BibitemShut {NoStop}%
\bibitem [{\citenamefont {Schiller}(2007)}]{schiller_hydrogenlike_2007}%
  \BibitemOpen
  \bibfield  {author} {\bibinfo {author} {\bibfnamefont {S.}~\bibnamefont
  {Schiller}},\ }\href {\doibase 10.1103/PhysRevLett.98.180801} {\bibfield
  {journal} {\bibinfo  {journal} {Phys. Rev. Lett.}\ }\textbf {\bibinfo
  {volume} {98}},\ \bibinfo {pages} {180801} (\bibinfo {year}
  {2007})}\BibitemShut {NoStop}%
\bibitem [{\citenamefont {Berengut}\ \emph {et~al.}(2010)\citenamefont
  {Berengut}, \citenamefont {Dzuba},\ and\ \citenamefont
  {Flambaum}}]{BerDzuFla10}%
  \BibitemOpen
  \bibfield  {author} {\bibinfo {author} {\bibfnamefont {J.~C.}\ \bibnamefont
  {Berengut}}, \bibinfo {author} {\bibfnamefont {V.~A.}\ \bibnamefont {Dzuba}},
  \ and\ \bibinfo {author} {\bibfnamefont {V.~V.}\ \bibnamefont {Flambaum}},\
  }\href {\doibase 10.1103/PhysRevLett.105.120801} {\bibfield  {journal}
  {\bibinfo  {journal} {Phys. Rev. Lett.}\ }\textbf {\bibinfo {volume} {105}},\
  \bibinfo {pages} {120801} (\bibinfo {year} {2010})}\BibitemShut {NoStop}%
\bibitem [{\citenamefont {Derevianko}\ \emph {et~al.}(2012)\citenamefont
  {Derevianko}, \citenamefont {Dzuba},\ and\ \citenamefont
  {Flambaum}}]{DerDzuFla12}%
  \BibitemOpen
  \bibfield  {author} {\bibinfo {author} {\bibfnamefont {A.}~\bibnamefont
  {Derevianko}}, \bibinfo {author} {\bibfnamefont {V.~A.}\ \bibnamefont
  {Dzuba}}, \ and\ \bibinfo {author} {\bibfnamefont {V.~V.}\ \bibnamefont
  {Flambaum}},\ }\href {\doibase 10.1103/PhysRevLett.109.180801} {\bibfield
  {journal} {\bibinfo  {journal} {Phys. Rev. Lett.}\ }\textbf {\bibinfo
  {volume} {109}},\ \bibinfo {pages} {180801} (\bibinfo {year}
  {2012})}\BibitemShut {NoStop}%
\bibitem [{\citenamefont {Dzuba}\ \emph {et~al.}(2012)\citenamefont {Dzuba},
  \citenamefont {Derevianko},\ and\ \citenamefont {Flambaum}}]{DzuDerFla12}%
  \BibitemOpen
  \bibfield  {author} {\bibinfo {author} {\bibfnamefont {V.~A.}\ \bibnamefont
  {Dzuba}}, \bibinfo {author} {\bibfnamefont {A.}~\bibnamefont {Derevianko}}, \
  and\ \bibinfo {author} {\bibfnamefont {V.~V.}\ \bibnamefont {Flambaum}},\
  }\href {\doibase 10.1103/PhysRevA.86.054501} {\bibfield  {journal} {\bibinfo
  {journal} {Phys. Rev. A}\ }\textbf {\bibinfo {volume} {86}},\ \bibinfo
  {pages} {054501} (\bibinfo {year} {2012})}\BibitemShut {NoStop}%
\bibitem [{\citenamefont {Dzuba}\ \emph {et~al.}(2013)\citenamefont {Dzuba},
  \citenamefont {Derevianko},\ and\ \citenamefont {Flambaum}}]{DzuDerFla13}%
  \BibitemOpen
  \bibfield  {author} {\bibinfo {author} {\bibfnamefont {V.~A.}\ \bibnamefont
  {Dzuba}}, \bibinfo {author} {\bibfnamefont {A.}~\bibnamefont {Derevianko}}, \
  and\ \bibinfo {author} {\bibfnamefont {V.~V.}\ \bibnamefont {Flambaum}},\
  }\href {\doibase 10.1103/PhysRevA.87.029906} {\bibfield  {journal} {\bibinfo
  {journal} {Phys. Rev. A}\ }\textbf {\bibinfo {volume} {87}},\ \bibinfo
  {pages} {029906} (\bibinfo {year} {2013})}\BibitemShut {NoStop}%
\bibitem [{\citenamefont {Derevianko}\ and\ \citenamefont
  {Pospelov}(2014)}]{DerPos14}%
  \BibitemOpen
  \bibfield  {author} {\bibinfo {author} {\bibfnamefont {A.}~\bibnamefont
  {Derevianko}}\ and\ \bibinfo {author} {\bibfnamefont {M.}~\bibnamefont
  {Pospelov}},\ }\href@noop {} {\bibfield  {journal} {\bibinfo  {journal}
  {Nature Phys.}\ }\textbf {\bibinfo {volume} {10}},\ \bibinfo {pages} {933}
  (\bibinfo {year} {2014})}\BibitemShut {NoStop}%
\bibitem [{\citenamefont {Stadnik}\ and\ \citenamefont
  {Flambaum}(2015)}]{StaFla15}%
  \BibitemOpen
  \bibfield  {author} {\bibinfo {author} {\bibfnamefont {Y.~V.}\ \bibnamefont
  {Stadnik}}\ and\ \bibinfo {author} {\bibfnamefont {V.~V.}\ \bibnamefont
  {Flambaum}},\ }\href {\doibase 10.1103/PhysRevLett.115.201301} {\bibfield
  {journal} {\bibinfo  {journal} {Phys. Rev. Lett.}\ }\textbf {\bibinfo
  {volume} {115}},\ \bibinfo {pages} {201301} (\bibinfo {year}
  {2015})}\BibitemShut {NoStop}%
\bibitem [{\citenamefont {Stadnik}\ and\ \citenamefont
  {Flambaum}(2016)}]{StaFla16}%
  \BibitemOpen
  \bibfield  {author} {\bibinfo {author} {\bibfnamefont {Y.~V.}\ \bibnamefont
  {Stadnik}}\ and\ \bibinfo {author} {\bibfnamefont {V.~V.}\ \bibnamefont
  {Flambaum}},\ }\href {\doibase 10.1103/PhysRevA.94.022111} {\bibfield
  {journal} {\bibinfo  {journal} {Phys. Rev. A}\ }\textbf {\bibinfo {volume}
  {94}},\ \bibinfo {pages} {022111} (\bibinfo {year} {2016})}\BibitemShut
  {NoStop}%
\bibitem [{\citenamefont {Schm\"oger}\ \emph {et~al.}(2015)\citenamefont
  {Schm\"oger}, \citenamefont {Schwarz}, \citenamefont {Baumann}, \citenamefont
  {Versolato}, \citenamefont {Piest}, \citenamefont {Pfeifer}, \citenamefont
  {Ullrich}, \citenamefont {Schmidt},\ and\ \citenamefont {{Crespo
  L{\'o}pez-Urrutia}}}]{schmoger_deceleration_2015}%
  \BibitemOpen
  \bibfield  {author} {\bibinfo {author} {\bibfnamefont {L.}~\bibnamefont
  {Schm\"oger}}, \bibinfo {author} {\bibfnamefont {M.}~\bibnamefont {Schwarz}},
  \bibinfo {author} {\bibfnamefont {T.~M.}\ \bibnamefont {Baumann}}, \bibinfo
  {author} {\bibfnamefont {O.~O.}\ \bibnamefont {Versolato}}, \bibinfo {author}
  {\bibfnamefont {B.}~\bibnamefont {Piest}}, \bibinfo {author} {\bibfnamefont
  {T.}~\bibnamefont {Pfeifer}}, \bibinfo {author} {\bibfnamefont
  {J.}~\bibnamefont {Ullrich}}, \bibinfo {author} {\bibfnamefont {P.~O.}\
  \bibnamefont {Schmidt}}, \ and\ \bibinfo {author} {\bibfnamefont {J.~R.}\
  \bibnamefont {{Crespo L{\'o}pez-Urrutia}}},\ }\href {\doibase
  10.1063/1.4934245} {\bibfield  {journal} {\bibinfo  {journal} {Rev. Sci.
  Instrum.}\ }\textbf {\bibinfo {volume} {86}},\ \bibinfo {pages} {103111}
  (\bibinfo {year} {2015})}\BibitemShut {NoStop}%
\bibitem [{\citenamefont {Schwarz}\ \emph {et~al.}(2012)\citenamefont
  {Schwarz}, \citenamefont {Versolato}, \citenamefont {Windberger},
  \citenamefont {Brunner}, \citenamefont {Ballance}, \citenamefont {Eberle},
  \citenamefont {Ullrich}, \citenamefont {Schmidt}, \citenamefont {Hansen},
  \citenamefont {Gingell}, \citenamefont {Drewsen},\ and\ \citenamefont
  {{Crespo L{\'o}pez-Urrutia}}}]{schwarz_cryogenic_2012}%
  \BibitemOpen
  \bibfield  {author} {\bibinfo {author} {\bibfnamefont {M.}~\bibnamefont
  {Schwarz}}, \bibinfo {author} {\bibfnamefont {O.~O.}\ \bibnamefont
  {Versolato}}, \bibinfo {author} {\bibfnamefont {A.}~\bibnamefont
  {Windberger}}, \bibinfo {author} {\bibfnamefont {F.~R.}\ \bibnamefont
  {Brunner}}, \bibinfo {author} {\bibfnamefont {T.}~\bibnamefont {Ballance}},
  \bibinfo {author} {\bibfnamefont {S.~N.}\ \bibnamefont {Eberle}}, \bibinfo
  {author} {\bibfnamefont {J.}~\bibnamefont {Ullrich}}, \bibinfo {author}
  {\bibfnamefont {P.~O.}\ \bibnamefont {Schmidt}}, \bibinfo {author}
  {\bibfnamefont {A.~K.}\ \bibnamefont {Hansen}}, \bibinfo {author}
  {\bibfnamefont {A.~D.}\ \bibnamefont {Gingell}}, \bibinfo {author}
  {\bibfnamefont {M.}~\bibnamefont {Drewsen}}, \ and\ \bibinfo {author}
  {\bibfnamefont {J.~R.}\ \bibnamefont {{Crespo L{\'o}pez-Urrutia}}},\ }\href
  {\doibase doi:10.1063/1.4742770} {\bibfield  {journal} {\bibinfo  {journal}
  {Rev. Sci. Instrum.}\ }\textbf {\bibinfo {volume} {83}},\ \bibinfo {pages}
  {083115} (\bibinfo {year} {2012})}\BibitemShut {NoStop}%
\bibitem [{\citenamefont {Leopold}\ \emph {et~al.}(2019)\citenamefont
  {Leopold}, \citenamefont {King}, \citenamefont {Micke}, \citenamefont
  {{Bautista-Salvador}}, \citenamefont {Heip}, \citenamefont {Ospelkaus},
  \citenamefont {{Crespo L{\'o}pez-Urrutia}},\ and\ \citenamefont
  {Schmidt}}]{leopold_cryogenic_2019}%
  \BibitemOpen
  \bibfield  {author} {\bibinfo {author} {\bibfnamefont {T.}~\bibnamefont
  {Leopold}}, \bibinfo {author} {\bibfnamefont {S.~A.}\ \bibnamefont {King}},
  \bibinfo {author} {\bibfnamefont {P.}~\bibnamefont {Micke}}, \bibinfo
  {author} {\bibfnamefont {A.}~\bibnamefont {{Bautista-Salvador}}}, \bibinfo
  {author} {\bibfnamefont {J.~C.}\ \bibnamefont {Heip}}, \bibinfo {author}
  {\bibfnamefont {C.}~\bibnamefont {Ospelkaus}}, \bibinfo {author}
  {\bibfnamefont {J.~R.}\ \bibnamefont {{Crespo L{\'o}pez-Urrutia}}}, \ and\
  \bibinfo {author} {\bibfnamefont {P.~O.}\ \bibnamefont {Schmidt}},\
  }\href@noop {} {\bibfield  {journal} {\bibinfo  {journal} {Rev. Sci.
  Instrum.}\ }\textbf {\bibinfo {volume} {90}},\ \bibinfo {pages} {073201}
  (\bibinfo {year} {2019})}\BibitemShut {NoStop}%
\bibitem [{\citenamefont {Micke}\ \emph {et~al.}(2019)\citenamefont {Micke},
  \citenamefont {Stark}, \citenamefont {King}, \citenamefont {Leopold},
  \citenamefont {Pfeifer}, \citenamefont {Schm\"oger}, \citenamefont {Schwarz},
  \citenamefont {Spie\ss}, \citenamefont {Schmidt},\ and\ \citenamefont
  {{Crespo L{\'o}pez-Urrutia}}}]{micke_closed-cycle_2019}%
  \BibitemOpen
  \bibfield  {author} {\bibinfo {author} {\bibfnamefont {P.}~\bibnamefont
  {Micke}}, \bibinfo {author} {\bibfnamefont {J.}~\bibnamefont {Stark}},
  \bibinfo {author} {\bibfnamefont {S.~A.}\ \bibnamefont {King}}, \bibinfo
  {author} {\bibfnamefont {T.}~\bibnamefont {Leopold}}, \bibinfo {author}
  {\bibfnamefont {T.}~\bibnamefont {Pfeifer}}, \bibinfo {author} {\bibfnamefont
  {L.}~\bibnamefont {Schm\"oger}}, \bibinfo {author} {\bibfnamefont
  {M.}~\bibnamefont {Schwarz}}, \bibinfo {author} {\bibfnamefont {L.~J.}\
  \bibnamefont {Spie\ss}}, \bibinfo {author} {\bibfnamefont {P.~O.}\
  \bibnamefont {Schmidt}}, \ and\ \bibinfo {author} {\bibfnamefont {J.~R.}\
  \bibnamefont {{Crespo L{\'o}pez-Urrutia}}},\ }\href@noop {} {\bibfield
  {journal} {\bibinfo  {journal} {Rev. Sci. Instrum.}\ }\textbf {\bibinfo
  {volume} {90}},\ \bibinfo {pages} {065104} (\bibinfo {year}
  {2019})}\BibitemShut {NoStop}%
\bibitem [{\citenamefont {{Berengut}}\ \emph {et~al.}(2012)\citenamefont
  {{Berengut}}, \citenamefont {{Dzuba}}, \citenamefont {{Flambaum}},\ and\
  \citenamefont {{Ong}}}]{BerDzuFla12}%
  \BibitemOpen
  \bibfield  {author} {\bibinfo {author} {\bibfnamefont {J.~C.}\ \bibnamefont
  {{Berengut}}}, \bibinfo {author} {\bibfnamefont {V.~A.}\ \bibnamefont
  {{Dzuba}}}, \bibinfo {author} {\bibfnamefont {V.~V.}\ \bibnamefont
  {{Flambaum}}}, \ and\ \bibinfo {author} {\bibfnamefont {A.}~\bibnamefont
  {{Ong}}},\ }\href@noop {} {\bibfield  {journal} {\bibinfo  {journal} {Phys.
  Rev. Lett.}\ }\textbf {\bibinfo {volume} {109}},\ \bibinfo {pages} {070802}
  (\bibinfo {year} {2012})}\BibitemShut {NoStop}%
\bibitem [{\citenamefont {Dzuba}\ \emph {et~al.}(2015)\citenamefont {Dzuba},
  \citenamefont {Safronova}, \citenamefont {Safronova},\ and\ \citenamefont
  {Flambaum}}]{DzuSafSaf15}%
  \BibitemOpen
  \bibfield  {author} {\bibinfo {author} {\bibfnamefont {V.~A.}\ \bibnamefont
  {Dzuba}}, \bibinfo {author} {\bibfnamefont {M.~S.}\ \bibnamefont
  {Safronova}}, \bibinfo {author} {\bibfnamefont {U.~I.}\ \bibnamefont
  {Safronova}}, \ and\ \bibinfo {author} {\bibfnamefont {V.~V.}\ \bibnamefont
  {Flambaum}},\ }\href@noop {} {\bibfield  {journal} {\bibinfo  {journal}
  {Phys. Rev. A}\ }\textbf {\bibinfo {volume} {92}},\ \bibinfo {pages} {060502}
  (\bibinfo {year} {2015})}\BibitemShut {NoStop}%
\bibitem [{\citenamefont {Kozlov}\ \emph {et~al.}(1996)\citenamefont {Kozlov},
  \citenamefont {Porsev},\ and\ \citenamefont {Flambaum}}]{KozPorFla96}%
  \BibitemOpen
  \bibfield  {author} {\bibinfo {author} {\bibfnamefont {M.~G.}\ \bibnamefont
  {Kozlov}}, \bibinfo {author} {\bibfnamefont {S.~G.}\ \bibnamefont {Porsev}},
  \ and\ \bibinfo {author} {\bibfnamefont {V.~V.}\ \bibnamefont {Flambaum}},\
  }\href@noop {} {\bibfield  {journal} {\bibinfo  {journal} {J. \ Phys. \ B}\
  }\textbf {\bibinfo {volume} {29}},\ \bibinfo {pages} {689} (\bibinfo {year}
  {1996})}\BibitemShut {NoStop}%
\bibitem [{\citenamefont {Shabaev}\ \emph {et~al.}(2013)\citenamefont
  {Shabaev}, \citenamefont {Tupitsyn},\ and\ \citenamefont
  {Yerokhin}}]{ShaTupYer13}%
  \BibitemOpen
  \bibfield  {author} {\bibinfo {author} {\bibfnamefont {V.~M.}\ \bibnamefont
  {Shabaev}}, \bibinfo {author} {\bibfnamefont {I.~I.}\ \bibnamefont
  {Tupitsyn}}, \ and\ \bibinfo {author} {\bibfnamefont {V.~A.}\ \bibnamefont
  {Yerokhin}},\ }\href {\doibase 10.1103/PhysRevA.88.012513} {\bibfield
  {journal} {\bibinfo  {journal} {Phys. Rev. A}\ }\textbf {\bibinfo {volume}
  {88}},\ \bibinfo {pages} {012513} (\bibinfo {year} {2013})}\BibitemShut
  {NoStop}%
\bibitem [{\citenamefont {Tupitsyn}\ \emph {et~al.}(2016)\citenamefont
  {Tupitsyn}, \citenamefont {Kozlov}, \citenamefont {Safronova}, \citenamefont
  {Shabaev},\ and\ \citenamefont {Dzuba}}]{TupKozSaf16}%
  \BibitemOpen
  \bibfield  {author} {\bibinfo {author} {\bibfnamefont {I.~I.}\ \bibnamefont
  {Tupitsyn}}, \bibinfo {author} {\bibfnamefont {M.~G.}\ \bibnamefont
  {Kozlov}}, \bibinfo {author} {\bibfnamefont {M.~S.}\ \bibnamefont
  {Safronova}}, \bibinfo {author} {\bibfnamefont {V.~M.}\ \bibnamefont
  {Shabaev}}, \ and\ \bibinfo {author} {\bibfnamefont {V.~A.}\ \bibnamefont
  {Dzuba}},\ }\href@noop {} {\bibfield  {journal} {\bibinfo  {journal} {Phys.
  Rev. Lett.}\ }\textbf {\bibinfo {volume} {117}},\ \bibinfo {pages} {253001}
  (\bibinfo {year} {2016})}\BibitemShut {NoStop}%
\bibitem [{\citenamefont {Dzuba}\ \emph {et~al.}(1996)\citenamefont {Dzuba},
  \citenamefont {Flambaum},\ and\ \citenamefont {Kozlov}}]{DzuFlaKoz96}%
  \BibitemOpen
  \bibfield  {author} {\bibinfo {author} {\bibfnamefont {V.~A.}\ \bibnamefont
  {Dzuba}}, \bibinfo {author} {\bibfnamefont {V.~V.}\ \bibnamefont {Flambaum}},
  \ and\ \bibinfo {author} {\bibfnamefont {M.~G.}\ \bibnamefont {Kozlov}},\
  }\href@noop {} {\bibfield  {journal} {\bibinfo  {journal} {Phys.\ Rev.\ A}\
  }\textbf {\bibinfo {volume} {54}},\ \bibinfo {pages} {3948} (\bibinfo {year}
  {1996})}\BibitemShut {NoStop}%
\bibitem [{\citenamefont {{Safronova}}\ \emph {et~al.}(2009)\citenamefont
  {{Safronova}}, \citenamefont {{Kozlov}}, \citenamefont {{Johnson}},\ and\
  \citenamefont {{Jiang}}}]{SafKozJoh09}%
  \BibitemOpen
  \bibfield  {author} {\bibinfo {author} {\bibfnamefont {M.~S.}\ \bibnamefont
  {{Safronova}}}, \bibinfo {author} {\bibfnamefont {M.~G.}\ \bibnamefont
  {{Kozlov}}}, \bibinfo {author} {\bibfnamefont {W.~R.}\ \bibnamefont
  {{Johnson}}}, \ and\ \bibinfo {author} {\bibfnamefont {D.}~\bibnamefont
  {{Jiang}}},\ }\href@noop {} {\bibfield  {journal} {\bibinfo  {journal} {Phys.
  Rev. A}\ }\textbf {\bibinfo {volume} {80}},\ \bibinfo {eid} {012516}
  (\bibinfo {year} {2009})}\BibitemShut {NoStop}%
\bibitem [{\citenamefont {Kozlov}(2004)}]{Koz04}%
  \BibitemOpen
  \bibfield  {author} {\bibinfo {author} {\bibfnamefont {M.~G.}\ \bibnamefont
  {Kozlov}},\ }\href@noop {} {\bibfield  {journal} {\bibinfo  {journal} {Int.
  J. Quant. Chem.}\ }\textbf {\bibinfo {volume} {100}},\ \bibinfo {pages} {336}
  (\bibinfo {year} {2004})}\BibitemShut {NoStop}%
\bibitem [{\citenamefont {Porsev}\ and\ \citenamefont
  {Derevianko}(2006)}]{PorDer06}%
  \BibitemOpen
  \bibfield  {author} {\bibinfo {author} {\bibfnamefont {S.~G.}\ \bibnamefont
  {Porsev}}\ and\ \bibinfo {author} {\bibfnamefont {A.}~\bibnamefont
  {Derevianko}},\ }\href@noop {} {\bibfield  {journal} {\bibinfo  {journal}
  {Phys. Rev. A}\ }\textbf {\bibinfo {volume} {73}},\ \bibinfo {pages} {012501}
  (\bibinfo {year} {2006})}\BibitemShut {NoStop}%
\bibitem [{\citenamefont {Safronova}\ \emph
  {et~al.}(2014{\natexlab{a}})\citenamefont {Safronova}, \citenamefont {Dzuba},
  \citenamefont {Flambaum}, \citenamefont {Safronova}, \citenamefont {Porsev},\
  and\ \citenamefont {Kozlov}}]{SafDzuFla14PRA1}%
  \BibitemOpen
  \bibfield  {author} {\bibinfo {author} {\bibfnamefont {M.~S.}\ \bibnamefont
  {Safronova}}, \bibinfo {author} {\bibfnamefont {V.~A.}\ \bibnamefont
  {Dzuba}}, \bibinfo {author} {\bibfnamefont {V.~V.}\ \bibnamefont {Flambaum}},
  \bibinfo {author} {\bibfnamefont {U.~I.}\ \bibnamefont {Safronova}}, \bibinfo
  {author} {\bibfnamefont {S.~G.}\ \bibnamefont {Porsev}}, \ and\ \bibinfo
  {author} {\bibfnamefont {M.~G.}\ \bibnamefont {Kozlov}},\ }\href@noop {}
  {\bibfield  {journal} {\bibinfo  {journal} {Phys. Rev. A}\ }\textbf {\bibinfo
  {volume} {90}},\ \bibinfo {pages} {042513} (\bibinfo {year}
  {2014}{\natexlab{a}})}\BibitemShut {NoStop}%
\bibitem [{\citenamefont {Safronova}\ \emph
  {et~al.}(2014{\natexlab{b}})\citenamefont {Safronova}, \citenamefont {Dzuba},
  \citenamefont {Flambaum}, \citenamefont {Safronova}, \citenamefont {Porsev},\
  and\ \citenamefont {Kozlov}}]{SafDzuFla14PRA2}%
  \BibitemOpen
  \bibfield  {author} {\bibinfo {author} {\bibfnamefont {M.~S.}\ \bibnamefont
  {Safronova}}, \bibinfo {author} {\bibfnamefont {V.~A.}\ \bibnamefont
  {Dzuba}}, \bibinfo {author} {\bibfnamefont {V.~V.}\ \bibnamefont {Flambaum}},
  \bibinfo {author} {\bibfnamefont {U.~I.}\ \bibnamefont {Safronova}}, \bibinfo
  {author} {\bibfnamefont {S.~G.}\ \bibnamefont {Porsev}}, \ and\ \bibinfo
  {author} {\bibfnamefont {M.~G.}\ \bibnamefont {Kozlov}},\ }\href@noop {}
  {\bibfield  {journal} {\bibinfo  {journal} {Phys. Rev. A}\ }\textbf {\bibinfo
  {volume} {90}},\ \bibinfo {pages} {052509} (\bibinfo {year}
  {2014}{\natexlab{b}})}\BibitemShut {NoStop}%
\bibitem [{\citenamefont {Cannon}\ and\ \citenamefont
  {Derevianko}(2004)}]{CanDer04}%
  \BibitemOpen
  \bibfield  {author} {\bibinfo {author} {\bibfnamefont {C.~C.}\ \bibnamefont
  {Cannon}}\ and\ \bibinfo {author} {\bibfnamefont {A.}~\bibnamefont
  {Derevianko}},\ }\href@noop {} {\bibfield  {journal} {\bibinfo  {journal}
  {Phys. Rev. A}\ }\textbf {\bibinfo {volume} {69}},\ \bibinfo {pages} {030502}
  (\bibinfo {year} {2004})}\BibitemShut {NoStop}%
\bibitem [{\citenamefont {Derevianko}\ and\ \citenamefont
  {Porsev}(2005)}]{DerPor05}%
  \BibitemOpen
  \bibfield  {author} {\bibinfo {author} {\bibfnamefont {A.}~\bibnamefont
  {Derevianko}}\ and\ \bibinfo {author} {\bibfnamefont {S.~G.}\ \bibnamefont
  {Porsev}},\ }\href@noop {} {\bibfield  {journal} {\bibinfo  {journal} {Phys.
  Rev. A}\ }\textbf {\bibinfo {volume} {71}},\ \bibinfo {pages} {032509}
  (\bibinfo {year} {2005})}\BibitemShut {NoStop}%
\bibitem [{\citenamefont {Ramsey}(1956)}]{Ram56}%
  \BibitemOpen
  \bibfield  {author} {\bibinfo {author} {\bibfnamefont {N.~F.}\ \bibnamefont
  {Ramsey}},\ }\href@noop {} {\emph {\bibinfo {title} {Molecular Beams}}}\
  (\bibinfo  {publisher} {Oxford Univ. Press},\ \bibinfo {address} {London},\
  \bibinfo {year} {1956})\BibitemShut {NoStop}%
\bibitem [{\citenamefont {Itano}(2000)}]{Ita00}%
  \BibitemOpen
  \bibfield  {author} {\bibinfo {author} {\bibfnamefont {W.}~\bibnamefont
  {Itano}},\ }\href@noop {} {\bibfield  {journal} {\bibinfo  {journal} {J. Res.
  Natl. Inst. Stand. Technol.}\ }\textbf {\bibinfo {volume} {105}},\ \bibinfo
  {pages} {829} (\bibinfo {year} {2000})}\BibitemShut {NoStop}%
\bibitem [{\citenamefont {Dub\'e}\ \emph {et~al.}(2005)\citenamefont {Dub\'e},
  \citenamefont {Madej}, \citenamefont {Bernard}, \citenamefont {Marmet},
  \citenamefont {Boulanger},\ and\ \citenamefont {Cundy}}]{dube_electric_2005}%
  \BibitemOpen
  \bibfield  {author} {\bibinfo {author} {\bibfnamefont {P.}~\bibnamefont
  {Dub\'e}}, \bibinfo {author} {\bibfnamefont {A.}~\bibnamefont {Madej}},
  \bibinfo {author} {\bibfnamefont {J.}~\bibnamefont {Bernard}}, \bibinfo
  {author} {\bibfnamefont {L.}~\bibnamefont {Marmet}}, \bibinfo {author}
  {\bibfnamefont {J.-S.}\ \bibnamefont {Boulanger}}, \ and\ \bibinfo {author}
  {\bibfnamefont {S.}~\bibnamefont {Cundy}},\ }\href {\doibase
  10.1103/PhysRevLett.95.033001} {\bibfield  {journal} {\bibinfo  {journal}
  {Phys. Rev. Lett.}\ }\textbf {\bibinfo {volume} {95}},\ \bibinfo {pages}
  {033001} (\bibinfo {year} {2005})}\BibitemShut {NoStop}%
\bibitem [{\citenamefont {Margolis}\ \emph {et~al.}(2004)\citenamefont
  {Margolis}, \citenamefont {Barwood}, \citenamefont {Huang}, \citenamefont
  {Klein}, \citenamefont {Lea}, \citenamefont {Szymaniec},\ and\ \citenamefont
  {Gill}}]{margolis_hertz-level_2004}%
  \BibitemOpen
  \bibfield  {author} {\bibinfo {author} {\bibfnamefont {H.~S.}\ \bibnamefont
  {Margolis}}, \bibinfo {author} {\bibfnamefont {G.~P.}\ \bibnamefont
  {Barwood}}, \bibinfo {author} {\bibfnamefont {G.}~\bibnamefont {Huang}},
  \bibinfo {author} {\bibfnamefont {H.~A.}\ \bibnamefont {Klein}}, \bibinfo
  {author} {\bibfnamefont {S.~N.}\ \bibnamefont {Lea}}, \bibinfo {author}
  {\bibfnamefont {K.}~\bibnamefont {Szymaniec}}, \ and\ \bibinfo {author}
  {\bibfnamefont {P.}~\bibnamefont {Gill}},\ }\href {\doibase
  10.1126/science.1105497} {\bibfield  {journal} {\bibinfo  {journal}
  {Science}\ }\textbf {\bibinfo {volume} {306}},\ \bibinfo {pages} {1355 }
  (\bibinfo {year} {2004})}\BibitemShut {NoStop}%
\bibitem [{\citenamefont {Chwalla}\ \emph {et~al.}(2009)\citenamefont
  {Chwalla}, \citenamefont {Benhelm}, \citenamefont {Kim}, \citenamefont
  {Kirchmair}, \citenamefont {Monz}, \citenamefont {Riebe}, \citenamefont
  {Schindler}, \citenamefont {Villar}, \citenamefont {H\"ansel}, \citenamefont
  {Roos}, \citenamefont {Blatt}, \citenamefont {Abgrall}, \citenamefont
  {Santarelli}, \citenamefont {Rovera},\ and\ \citenamefont
  {Laurent}}]{chwalla_absolute_2009}%
  \BibitemOpen
  \bibfield  {author} {\bibinfo {author} {\bibfnamefont {M.}~\bibnamefont
  {Chwalla}}, \bibinfo {author} {\bibfnamefont {J.}~\bibnamefont {Benhelm}},
  \bibinfo {author} {\bibfnamefont {K.}~\bibnamefont {Kim}}, \bibinfo {author}
  {\bibfnamefont {G.}~\bibnamefont {Kirchmair}}, \bibinfo {author}
  {\bibfnamefont {T.}~\bibnamefont {Monz}}, \bibinfo {author} {\bibfnamefont
  {M.}~\bibnamefont {Riebe}}, \bibinfo {author} {\bibfnamefont
  {P.}~\bibnamefont {Schindler}}, \bibinfo {author} {\bibfnamefont
  {A.}~\bibnamefont {Villar}}, \bibinfo {author} {\bibfnamefont
  {W.}~\bibnamefont {H\"ansel}}, \bibinfo {author} {\bibfnamefont
  {C.}~\bibnamefont {Roos}}, \bibinfo {author} {\bibfnamefont {R.}~\bibnamefont
  {Blatt}}, \bibinfo {author} {\bibfnamefont {M.}~\bibnamefont {Abgrall}},
  \bibinfo {author} {\bibfnamefont {G.}~\bibnamefont {Santarelli}}, \bibinfo
  {author} {\bibfnamefont {G.}~\bibnamefont {Rovera}}, \ and\ \bibinfo {author}
  {\bibfnamefont {P.}~\bibnamefont {Laurent}},\ }\href {\doibase
  10.1103/PhysRevLett.102.023002} {\bibfield  {journal} {\bibinfo  {journal}
  {Phys. Rev. Lett.}\ }\textbf {\bibinfo {volume} {102}},\ \bibinfo {pages}
  {023002} (\bibinfo {year} {2009})}\BibitemShut {NoStop}%
\bibitem [{\citenamefont {{Madej}}\ \emph {et~al.}(2012)\citenamefont
  {{Madej}}, \citenamefont {{Dub{\'e}}}, \citenamefont {{Zhou}}, \citenamefont
  {{Bernard}},\ and\ \citenamefont {{Gertsvolf}}}]{MadDubZho12}%
  \BibitemOpen
  \bibfield  {author} {\bibinfo {author} {\bibfnamefont {A.~A.}\ \bibnamefont
  {{Madej}}}, \bibinfo {author} {\bibfnamefont {P.}~\bibnamefont {{Dub{\'e}}}},
  \bibinfo {author} {\bibfnamefont {Z.}~\bibnamefont {{Zhou}}}, \bibinfo
  {author} {\bibfnamefont {J.~E.}\ \bibnamefont {{Bernard}}}, \ and\ \bibinfo
  {author} {\bibfnamefont {M.}~\bibnamefont {{Gertsvolf}}},\ }\href {\doibase
  10.1103/PhysRevLett.109.203002} {\bibfield  {journal} {\bibinfo  {journal}
  {Phys. Rev. Lett.}\ }\textbf {\bibinfo {volume} {109}},\ \bibinfo {eid}
  {203002} (\bibinfo {year} {2012})}\BibitemShut {NoStop}%
\bibitem [{\citenamefont {{Dub{\'e}}}\ \emph {et~al.}(2013)\citenamefont
  {{Dub{\'e}}}, \citenamefont {{Madej}}, \citenamefont {{Zhou}},\ and\
  \citenamefont {{Bernard}}}]{DubMadZho13}%
  \BibitemOpen
  \bibfield  {author} {\bibinfo {author} {\bibfnamefont {P.}~\bibnamefont
  {{Dub{\'e}}}}, \bibinfo {author} {\bibfnamefont {A.~A.}\ \bibnamefont
  {{Madej}}}, \bibinfo {author} {\bibfnamefont {Z.}~\bibnamefont {{Zhou}}}, \
  and\ \bibinfo {author} {\bibfnamefont {J.~E.}\ \bibnamefont {{Bernard}}},\
  }\href {\doibase 10.1103/PhysRevA.87.023806} {\bibfield  {journal} {\bibinfo
  {journal} {\pra}\ }\textbf {\bibinfo {volume} {87}},\ \bibinfo {eid} {023806}
  (\bibinfo {year} {2013})}\BibitemShut {NoStop}%
\bibitem [{\citenamefont {Shaniv}\ \emph {et~al.}(2016)\citenamefont {Shaniv},
  \citenamefont {Akerman},\ and\ \citenamefont {Ozeri}}]{ShaAkeOze16}%
  \BibitemOpen
  \bibfield  {author} {\bibinfo {author} {\bibfnamefont {R.}~\bibnamefont
  {Shaniv}}, \bibinfo {author} {\bibfnamefont {N.}~\bibnamefont {Akerman}}, \
  and\ \bibinfo {author} {\bibfnamefont {R.}~\bibnamefont {Ozeri}},\
  }\href@noop {} {\bibfield  {journal} {\bibinfo  {journal} {Phys. Rev. Lett.}\
  }\textbf {\bibinfo {volume} {116}},\ \bibinfo {pages} {140801} (\bibinfo
  {year} {2016})}\BibitemShut {NoStop}%
\bibitem [{\citenamefont {{Kozlov}}\ and\ \citenamefont
  {{Porsev}}(1999)}]{KozPor99a}%
  \BibitemOpen
  \bibfield  {author} {\bibinfo {author} {\bibfnamefont {M.~G.}\ \bibnamefont
  {{Kozlov}}}\ and\ \bibinfo {author} {\bibfnamefont {S.~G.}\ \bibnamefont
  {{Porsev}}},\ }\href {\doibase 10.1007/s100530050229} {\bibfield  {journal}
  {\bibinfo  {journal} {Eur.~Phys.~J.~D}\ }\textbf {\bibinfo {volume} {5}},\
  \bibinfo {pages} {59} (\bibinfo {year} {1999})}\BibitemShut {NoStop}%
\bibitem [{\citenamefont {Safronova}\ \emph {et~al.}(1999)\citenamefont
  {Safronova}, \citenamefont {Johnson},\ and\ \citenamefont
  {Derevianko}}]{SafJohDer99}%
  \BibitemOpen
  \bibfield  {author} {\bibinfo {author} {\bibfnamefont {M.~S.}\ \bibnamefont
  {Safronova}}, \bibinfo {author} {\bibfnamefont {W.~R.}\ \bibnamefont
  {Johnson}}, \ and\ \bibinfo {author} {\bibfnamefont {A.}~\bibnamefont
  {Derevianko}},\ }\href@noop {} {\bibfield  {journal} {\bibinfo  {journal}
  {Phys. Rev. A}\ }\textbf {\bibinfo {volume} {60}},\ \bibinfo {pages} {4476}
  (\bibinfo {year} {1999})}\BibitemShut {NoStop}%
\bibitem [{\citenamefont {Berkeland}\ \emph {et~al.}(1998)\citenamefont
  {Berkeland}, \citenamefont {Miller}, \citenamefont {Bergquist}, \citenamefont
  {Itano},\ and\ \citenamefont {Wineland}}]{BerMilBer98}%
  \BibitemOpen
  \bibfield  {author} {\bibinfo {author} {\bibfnamefont {D.~J.}\ \bibnamefont
  {Berkeland}}, \bibinfo {author} {\bibfnamefont {J.~D.}\ \bibnamefont
  {Miller}}, \bibinfo {author} {\bibfnamefont {J.~C.}\ \bibnamefont
  {Bergquist}}, \bibinfo {author} {\bibfnamefont {W.~M.}\ \bibnamefont
  {Itano}}, \ and\ \bibinfo {author} {\bibfnamefont {D.~J.}\ \bibnamefont
  {Wineland}},\ }\href@noop {} {\bibfield  {journal} {\bibinfo  {journal} {J.
  Appl. Phys.}\ }\textbf {\bibinfo {volume} {83}},\ \bibinfo {pages} {5025}
  (\bibinfo {year} {1998})}\BibitemShut {NoStop}%
\bibitem [{\citenamefont {Keller}\ \emph {et~al.}(2015)\citenamefont {Keller},
  \citenamefont {Partner}, \citenamefont {Burgermeister},\ and\ \citenamefont
  {Mehlst\"{a}ubler}}]{KelParBur15}%
  \BibitemOpen
  \bibfield  {author} {\bibinfo {author} {\bibfnamefont {J.}~\bibnamefont
  {Keller}}, \bibinfo {author} {\bibfnamefont {H.~L.}\ \bibnamefont {Partner}},
  \bibinfo {author} {\bibfnamefont {T.}~\bibnamefont {Burgermeister}}, \ and\
  \bibinfo {author} {\bibfnamefont {T.~E.}\ \bibnamefont {Mehlst\"{a}ubler}},\
  }\href@noop {} {\bibfield  {journal} {\bibinfo  {journal} {J. Appl. Phys.}\
  }\textbf {\bibinfo {volume} {118}},\ \bibinfo {pages} {104501} (\bibinfo
  {year} {2015})}\BibitemShut {NoStop}%
\bibitem [{\citenamefont {Edelstein}\ and\ \citenamefont
  {Karraker}(1975)}]{EdeKar75}%
  \BibitemOpen
  \bibfield  {author} {\bibinfo {author} {\bibfnamefont {N.}~\bibnamefont
  {Edelstein}}\ and\ \bibinfo {author} {\bibfnamefont {D.~G.}\ \bibnamefont
  {Karraker}},\ }\href@noop {} {\bibfield  {journal} {\bibinfo  {journal} {J.
  Chem. Phys.}\ }\textbf {\bibinfo {volume} {62}},\ \bibinfo {pages} {938}
  (\bibinfo {year} {1975})}\BibitemShut {NoStop}%
\bibitem [{\citenamefont {Bernard}\ \emph {et~al.}(1998)\citenamefont
  {Bernard}, \citenamefont {Marmet},\ and\ \citenamefont
  {Madej}}]{bernard_laser_1998}%
  \BibitemOpen
  \bibfield  {author} {\bibinfo {author} {\bibfnamefont {J.~E.}\ \bibnamefont
  {Bernard}}, \bibinfo {author} {\bibfnamefont {L.}~\bibnamefont {Marmet}}, \
  and\ \bibinfo {author} {\bibfnamefont {A.~A.}\ \bibnamefont {Madej}},\ }\href
  {\doibase 10.1016/S0030-4018(98)00121-7} {\bibfield  {journal} {\bibinfo
  {journal} {Opt. Commun.}\ }\textbf {\bibinfo {volume} {150}},\ \bibinfo
  {pages} {170} (\bibinfo {year} {1998})}\BibitemShut {NoStop}%
\bibitem [{\citenamefont {Breit}\ and\ \citenamefont {Rabi}(1931)}]{BreRab31}%
  \BibitemOpen
  \bibfield  {author} {\bibinfo {author} {\bibfnamefont {G.}~\bibnamefont
  {Breit}}\ and\ \bibinfo {author} {\bibfnamefont {I.}~\bibnamefont {Rabi}},\
  }\href@noop {} {\bibfield  {journal} {\bibinfo  {journal} {Phys. Rev.}\
  }\textbf {\bibinfo {volume} {28}},\ \bibinfo {pages} {2082} (\bibinfo {year}
  {1931})}\BibitemShut {NoStop}%
\bibitem [{\citenamefont {Rosenband}\ \emph {et~al.}(2007)\citenamefont
  {Rosenband}, \citenamefont {Schmidt}, \citenamefont {Hume}, \citenamefont
  {Itano}, \citenamefont {Fortier}, \citenamefont {Stalnaker}, \citenamefont
  {Kim}, \citenamefont {Diddams}, \citenamefont {Koelemeij}, \citenamefont
  {Bergquist},\ and\ \citenamefont {Wineland}}]{rosenband_observation_2007}%
  \BibitemOpen
  \bibfield  {author} {\bibinfo {author} {\bibfnamefont {T.}~\bibnamefont
  {Rosenband}}, \bibinfo {author} {\bibfnamefont {P.~O.}\ \bibnamefont
  {Schmidt}}, \bibinfo {author} {\bibfnamefont {D.~B.}\ \bibnamefont {Hume}},
  \bibinfo {author} {\bibfnamefont {W.~M.}\ \bibnamefont {Itano}}, \bibinfo
  {author} {\bibfnamefont {T.~M.}\ \bibnamefont {Fortier}}, \bibinfo {author}
  {\bibfnamefont {J.~E.}\ \bibnamefont {Stalnaker}}, \bibinfo {author}
  {\bibfnamefont {K.}~\bibnamefont {Kim}}, \bibinfo {author} {\bibfnamefont
  {S.~A.}\ \bibnamefont {Diddams}}, \bibinfo {author} {\bibfnamefont
  {J.~C.~J.}\ \bibnamefont {Koelemeij}}, \bibinfo {author} {\bibfnamefont
  {J.~C.}\ \bibnamefont {Bergquist}}, \ and\ \bibinfo {author} {\bibfnamefont
  {D.~J.}\ \bibnamefont {Wineland}},\ }\href {\doibase
  10.1103/PhysRevLett.98.220801} {\bibfield  {journal} {\bibinfo  {journal}
  {Phys. Rev. Lett.}\ }\textbf {\bibinfo {volume} {98}},\ \bibinfo {pages}
  {220801} (\bibinfo {year} {2007})}\BibitemShut {NoStop}%
\bibitem [{\citenamefont {Wineland}\ \emph {et~al.}(1983)\citenamefont
  {Wineland}, \citenamefont {Bollinger},\ and\ \citenamefont
  {Itano}}]{wineland_laser-fluorescence_1983}%
  \BibitemOpen
  \bibfield  {author} {\bibinfo {author} {\bibfnamefont {D.~J.}\ \bibnamefont
  {Wineland}}, \bibinfo {author} {\bibfnamefont {J.~J.}\ \bibnamefont
  {Bollinger}}, \ and\ \bibinfo {author} {\bibfnamefont {W.~M.}\ \bibnamefont
  {Itano}},\ }\href {\doibase 10.1103/PhysRevLett.50.628} {\bibfield  {journal}
  {\bibinfo  {journal} {Phys. Rev. Lett.}\ }\textbf {\bibinfo {volume} {50}},\
  \bibinfo {pages} {628} (\bibinfo {year} {1983})}\BibitemShut {NoStop}%
\end{thebibliography}

\end{document}